\documentclass[preprint,12pt]{elsarticle}
\usepackage[english]{babel}
\usepackage{mathrsfs}
\usepackage{amsmath}
\usepackage{amssymb} 
\usepackage{graphicx}
\usepackage{siunitx}
\usepackage{caption}
\usepackage{subcaption}
\usepackage[percent]{overpic}
\usepackage{multirow}
\usepackage{mathrsfs}
 \usepackage{calligra}
\usepackage{cancel}
\usepackage{rotating}
\usepackage{tikz}
\usepackage{placeins}
\usepackage{lipsum}
\usepackage{mathtools}
\DeclarePairedDelimiter{\ceil}{\lceil}{\rceil}
\usepackage{listings}
\usepackage[colorlinks=true, allcolors=blue]{hyperref}

\NewDocumentCommand{\codeword}{v}{%
\texttt{\textcolor{blue}{#1}}%
}
\DeclareMathOperator{\Tr}{Tr}

\journal{journal X}

\begin{document}

\begin{frontmatter}

\title{Grid-agnostic volume of fluid approach with interface sharpening and surface tension  for compressible multiphase flows}
\author[1]{Joseph J. Marziale} 
\author[1]{Jason Sun}
\author[1]{David Salac}
\author[1]{James Chen}
\affiliation[1]{
            addressline={Department of Mechanical and Aerospace Engineering}, 
            organization={The State University of New York at Buffalo},
            city={Buffalo},
            state={NY},
            postcode={14260},
            country={United States}}
\ead{chenjm@buffalo.edu}

\begin{abstract} The interfacial diffusion associated with finite volume method (FVM) discretizations of multiphase flows creates the need for an interface sharpening mechanism. Such solutions for structured quadrilateral grids are well documented, but various engineering applications require mesh designs specific to the irregular geometry of the physical system it is modeling. Therefore this study casts interface sharpening as an antidiffusive volumetric body force whose calculation procedure is generalizable to an arbitrarily constructed grid. The force magnitude is derived at cell centers as a function of the local compressible flow characteristics and the geometry of the cell neighborhood. The flow model uses an AUSM+up based method for flux evaluation and imposes a stiffened equation of state onto each of the fluids in order to close the linear system and extract auxiliary variables. Validation tests show good agreement with the Young-Laplace condition whereby the interface converges to the analytical solution corresponding to a balance between a pressure jump and interfacial forces. Further results show the recovery of a circle starting from a shape with highly variational curvature through the combined effects of surface tension and interface sharpening. Lastly shear-driven droplet pinchoff results show good agreement with droplet shapes provided by the surrounding literature at various Weber-Ohnesorge number combinations. \end{abstract}

\end{frontmatter}

\color{black}
\section{Introduction}\label{sec-intro}


Discretization of differential equations, such as the Navier-Stokes equation, results in artificial numerical diffusion. For multiphase flows this diffusion results in a smearing of the interface over time. However, precise interfacial dynamics determine the strength of capillary effects, heat transfer rates,  and phase transitions \cite{gennes2004capillarity,eggers2024coalescence,hsu2012surface,kim2016effect,dendukuri2005controlled,tanaka2001interplay}. Therefore it is crucial to retain the interface in multiphase flows. One potential solution to interface smearing is the use of refined grids, as the numerical approximations of the derivatives have a smaller error. However, a finer grid results in a numerical system that involves a greater number of points, and thus the solution time increases~\cite{marziale2024cvi}. Adaptive mesh refinement can be used to reduce unnecessary computational cost, dynamically refining the region near the interface while coarsening the far field~\cite{berger1984adaptive, berger1989local, chen2014thickness, rodriguez2013parallel, ngo2017multi}. This reduces some computational overhead but comes at the cost of interpolation error and algorithm complexity of refinement. Level set methods and other interface tracking algorithms are simpler to implement, but are decoupled from the physically constrained flow system and do not respect mass conservation without an additional VOF-informed or marker particle-informed mass correction process \cite{salac2011level,salac2016general,zeng2023consistent,enright2002use}. An  alternative which is both simple and conservative is the recent development of interface sharpening schemes, which have increased in popularity in recent years~\cite{he2024immiscibility,li2020interface,nguyen2017volume,so2012anti}. Interface sharpening schemes resolve the interface of two immiscible flows that otherwise exhibit artificial numerical diffusion. Examples of engineering applications that use interface sharpening schemes include oceanic and coastal wave breaking~\cite{sambe2011numerical}; underwater explosion~\cite{zhao2022interface}; rocket propulsion~\cite{barbeau2024assessment}; direct ink writing~\cite{shah2023multi,sun2024icme,sun2023damage}; scramjet fuel ignition~\cite{talbot2020numerical}; the liquid-vapor dynamics of boiling, evaporation, and condensation~\cite{hardt2008evaporation,sato2018examples}; and liquid-liquid oil extraction~\cite{wardle2013hybrid}. 


Several interface sharpening methods are reviewed by Anderson~\cite{anderson1998diffuse}, among which is the popular method of introducing a numerically artificial sharpening term to the interface advection equation \cite{lamorgese2017modeling, soligo2019mass, li2024family, li2024high}. Typically the interface velocity is decomposed into its tangential and normal components, and the normal component is modified to induce interface compression \cite{sun2007sharp}. Modeling the normal velocity as proportional to the curvature first appears in work by Osher and Sethian \cite{sethian1999level,osher2004level}. From this a counter-term approach which enforces zero curvature-driven motion was proposed by Folch and Casademunt \cite{folch1999phase}. To control the interface thickness to which the sharpening term converges, Boettinger et al \cite{boettinger2002phase} represent the interface as a hyperbolic tangent kernel with a transition region length parameter. This work culminated into a negative diffusion method introduced by Chiu and Lin \cite{chiu2011conservative} which is both conservative and only requires computation of second derivatives. Recently this method was adapted to a second-order structured quadrilateral discretization scheme and critiqued through error analysis by Jain~\cite{jain2022accurate}. In addition, Mirjalili et al delineates the boundedness and stability regions of this method \cite{mirjalili2020conservative}.


Interface sharpening schemes such as those discussed are commonly discretized to fit a uniform quadrilateral grid. However, there is a desire to overlay an arbitrary grid onto a fluid system which stems from the inherent geometric complexity of many applications. Flow surrounding an aircraft acts in tandem with the detailed curvature and angles of the wings and fuselage~\cite{kyrkos2012assessment, bhadauria2021lattice}. Multiphase flow of oil, water, and gas through petroleum reservoirs involves irregular pore shapes and fractured rock formations~\cite{jackson2013reservoir}. The air-water interface of breaking waves interacts dynamically with coastal reefs, shorelines, and other natural topography~\cite{xie2015two,tsao2023high}. In all such studies, the fluid behavior is best captured when the grid is fitted around these local regions of interest. 

Literature that accommodates these applications by generalizing an interface sharpening method to fit an arbitrary grid has recently developed \cite{hwang2024robust,jiang2025enhanced,tonicello2024high} but historically has shown to be more challenging to produce and therefore scarce. Early geometric VOF approaches, including PLIC, achieved sharp interface tracking by calculating planar subvolumes within structured cells as a reconstructed approximation of the interface position \cite{rider1995stretching,rider1998reconstructing}. 
\textcolor{black}{The general solution to the subvolume of an arbitrary polyhedron sliced by a plane as well as the possibility of concavity for unstructured grid cells remain as an active research \cite{JCP2008_LH,JCP2016_DF,IJNMF2019_DT}. Several efforts have been extended into arbitrary polyhedra \cite{JCP2019_LHGF,CC2024_LH}.
Applications of these research to PLIC reconstruction on arbitrary grids have recently been published \cite{CC2022_LH,CMAME2024_L}.}
Algebraic methods such as THINC emerged to avoid these limitations, replacing geometric reconstruction with an analytic profile for the volume fraction which can be integrated across a cell face \cite{xiao2005simple}. Original THINC formulations were tailored to structured quads, but the subsequent THINC/QQ extension by Xie and Xiao employs Gaussian quadrature over the faces of an arbitrary cell structure to approximate the face flux as a weighted sum, permitting unstructured grid adaptation \cite{xie2017toward,chen2022revisit,chen2023accurate}. This successfully sharpens the interface on unstructured grids but can introduce spurious oscillations near sharp gradients, as is typical for higher order reconstruction problems. To address this, Wakimura et al computes both THINC/QQ and a MUSCL-type reconstruction within each cell and selects the one yielding smaller boundary variation \cite{wakimura2025low}. However, a multi-point quadrature calculation per face per time step introduces computational expense. In contrast, the recent algebraic method by Kim et al uses a logistic sigmoid as a slope limiter on the volume fraction field at faces to steepen its gradient, and this method is much simpler to implement at the sacrifice of some accuracy \cite{chen2022revisit,kim2021efficient,adebayo2025review}. Both Kim's and Wakimura's schemes operate as face-centered flux correction terms and require evaluation of higher order interface projections across the cell faces and thus the success of the methods can vary based on the local mesh quality. In contrast, cell-centered schemes avoid such projections but are more seldom explored, especially outside of an incompressibility assumption on the velocity field.

\color{black}

This study is dedicated to developing a model that computes interface sharpening and surface tension terms of multiphase compressible flows valid to an arbitrarily constructed cell center distribution. The structure of the rest of the study is as follows. Section \ref{sec-theory} presents the derivation and discretization of the interface sharpening and surface tension terms, and their relationship to the multiphase conserved system. Section \ref{sec-results} validates the proposed model through a series of test cases, followed by an analysis of droplet shape physics based on the Weber-Ohnesorge number combination that characterizes the flow. Section \ref{sec-conclusion} concludes the study.

\section{Mathematical model}\label{sec-theory}

This section introduces the compressible multiphase flow system including interface sharpening and surface tension terms (Sec. \ref{sec-mfs}) and presents these terms' discretization for an arbitrarily constructed neighborhood of control volumes (Sec. \ref{sec-gradcomp}). In addition, the model's processes of closing the conserved system via fluid equations of state (Sec. \ref{sec-closure}) and evaluating fluxes (Sec. \ref{sec-fluxeval})  are explained.

\subsection{Multiphase flow system}\label{sec-mfs}
Let $\rho,\vec{u},e,\phi$ be density, velocity, internal energy, and volume fraction respectively; let the superscripts $^{(I)},^{(II)}$ denote quantities of one of two phases; let $\phi=\phi^{(I)}=1-\phi^{(II)}$; and let $\vec{S}$ be a vector of source terms. Consider the numerical diffusion associated with typical finite discretizations of a multiphase Euler system where the partial density of one of two phases ($\rho^{(I)}\phi$), mass density $(\rho)$, momentum density $(\rho\vec{u})$, and energy density ($\rho e$) per unit volume are conserved:
\begin{equation}\label{eq-euler2}
    \frac{\partial }{\partial t}\begin{Bmatrix}
        \rho^{(I)}\phi\\\rho\\\rho\vec{u}\\\rho e
    \end{Bmatrix} + \nabla\cdot\begin{Bmatrix}
        \rho^{(I)}\phi\vec{u}\\\rho\vec{u}\\\rho\vec{u}\otimes\vec{u}+p\textbf{I}-\boldsymbol{\tau}\\\rho H \vec{u}-\boldsymbol{\tau}\cdot\vec{u}
    \end{Bmatrix}  = \vec{S},
\end{equation}
where $t$ is time, $\textbf{I}$ is the identity matrix, $\boldsymbol{\tau}=\mu(\nabla\vec{u} + \nabla\vec{u}^T) - \frac{2}{3}\mu (\nabla\cdot\vec{u})\textbf{I}$ is the fluid stress tensor, $\mu=\phi \mu^{(I)}+(1-\phi)\mu^{(II)}$ is a volume fraction-weighted mixture viscosity coefficient, and $\rho H=\rho e+p$. Numerical diffusion results from converting Eq.~\eqref{eq-euler2} to a weak form within a finite cell volume $V$ and truncating higher order terms off the Taylor series expansion of the fluxes. In particular, the volume fraction interface smears due to truncation error from the partial density equation. Integrate the first equation in the linear system with respect to $V$ and apply the divergence theorem to get
\begin{equation}
    \int_V \Bigl[\frac{\partial }{\partial t} (\rho^{(I)} \phi ) + \nabla\cdot(\rho^{(I)}\phi\vec{u})\Bigl]dV = \frac{\partial}{\partial t} \int_V\rho^{(I)}\phi dV + \oint_{S } (\rho^{(I)}\phi\vec{u}) \cdot \vec{n}  dS = 0,
\end{equation}
where $S$ refers to the cell volume's surface and $\vec{n}$ is the surface outward normal. The flux term $\rho^{(I)}\phi\vec{u} := \vec{P}$ is discretized as a sum of the cell volume's face  contributions, such that
\begin{equation}\label{eq-surfaceintapprox}
    \oint_S \vec{P}\cdot\vec{n}dS =   \sum_{f\in V}\vec{P}^{|f} \cdot \vec{n}_f
\end{equation}
where $f$ denotes a face and $\vec{n}_f$ is the face area vector. From here, the finite volume method introduces two types of error. 

($i$) First, the field variables are interpolated to the faces from the cell centers. The Taylor series expansion of the flux about the cell centroid $z$ is
\begin{multline}
    \vec{P}^{|f} = \vec{P}^{|z} + \nabla\vec{P}^{|z}\cdot \vec{\mathcal{X}} + \frac{1}{2}H[\vec{P}^{|z}]: \Bigl[\vec{\mathcal{X}}\otimes \vec{\mathcal{X}}\Bigl]  \\ + \sum_{n=3}^\infty \frac{1}{n!}\frac{\partial^n \vec{P}^{|z} }{\partial x_{\alpha_1}\partial x_{\alpha_2} \ \dotsc \  \partial x_{\alpha_n}} \mathcal{X}_{\alpha_1}\mathcal{X}_{\alpha_2}\dotsc \mathcal{X}_{\alpha_n} 
\end{multline}
where $\vec{\mathcal{X}} := \vec{x}^{|f}-\vec{x}^{|z}$, i.e. the vector connecting the face to the centroid; $H$ is a Hessian function;  and $\alpha_i$ are dummy indices for which Einstein summation 

is implied. Typically the face is linearly interpolated such that $\vec{P}^{|f}_{lin} = \vec{P}^{|z} + \nabla\vec{P}^{|z}\cdot \vec{\mathcal{X}}$, which introduces a componentwise truncation error of
\begin{equation}\label{eq-type1error}
    \vec{\mathcal{E}}_{\vec{P}^{|f}}:= \vec{P}^{|f} - \vec{P}^{|f}_{lin} = \mathcal{O}(H[\vec{P}^{|z}]: \Bigl[\vec{\mathcal{X}}\otimes \vec{\mathcal{X}}\Bigl]) \ \ ,
\end{equation}
where $\mathcal{O}$ describes orders of magnitude. 

($ii$) Second, the gradient of the flux term $\vec{P}$ at the center is approximated via truncation. If a cell center has $M$ cell-centered neighbors then the gradient is calculated via
\begin{multline}\label{eq-gradPapprox}
    \begin{bmatrix}
        \vec{\mathcal{Z}}^{(1)}\cdot \vec{e}_{x_1} & \vec{\mathcal{Z}}^{(1)}\cdot \vec{e}_{x_2} & \vec{\mathcal{Z}}^{(1)}\cdot \vec{e}_{x_3}\\
        \vec{\mathcal{Z}}^{(2)}\cdot \vec{e}_{x_1} & \vec{\mathcal{Z}}^{(2)}\cdot \vec{e}_{x_2} & \vec{\mathcal{Z}}^{(2)}\cdot \vec{e}_{x_3}\\
        \vdots \\
        \vec{\mathcal{Z}}^{(M)}\cdot \vec{e}_{x_1} & \vec{\mathcal{Z}}^{(M)}\cdot \vec{e}_{x_2} & \vec{\mathcal{Z}}^{(M)}\cdot \vec{e}_{x_3}\\
    \end{bmatrix}\nabla\vec{P}^{|z} = \begin{Bmatrix}
        \vec{P}^{|z_1} - \vec{P}^{|z}\\
        \vec{P}^{|z_2} - \vec{P}^{|z}\\
        \vdots \\
        \vec{P}^{|z_M} - \vec{P}^{|z}
    \end{Bmatrix} \\
    -\frac{1}{2}\begin{Bmatrix}
        (\vec{\mathcal{Z}}^{(1)}\cdot \vec{e}_{x_1})^2 & (\vec{\mathcal{Z}}^{(1)}\cdot \vec{e}_{x_1})(\vec{\mathcal{Z}}^{(1)}\cdot \vec{e}_{x_2}) &  & (\vec{\mathcal{Z}}^{(1)}\cdot \vec{e}_{x_3})^2\\
        (\vec{\mathcal{Z}}^{(2)}\cdot \vec{e}_{x_1})^2 & (\vec{\mathcal{Z}}^{(2)}\cdot \vec{e}_{x_1})(\vec{\mathcal{Z}}^{(2)}\cdot \vec{e}_{x_2}) & \dotsc  & (\vec{\mathcal{Z}}^{(2)}\cdot \vec{e}_{x_3})^2\\ & \vdots  \\ (\vec{\mathcal{Z}}^{(M)}\cdot \vec{e}_{x_1})^2 & (\vec{\mathcal{Z}}^{(M)}\cdot \vec{e}_{x_1})(\vec{\mathcal{Z}}^{(M)}\cdot \vec{e}_{x_2}) &  & (\vec{\mathcal{Z}}^{(M)}\cdot \vec{e}_{x_3})^2\\
    \end{Bmatrix} \begin{Bmatrix}
        \partial_{x_1}^2 \vec{P}^{|z}\\
        \partial_{x_1}\partial_{x_2}  \vec{P}^{|z}\\
        \vdots \\
        \partial_{x_3}^2 \vec{P}^{|z}
    \end{Bmatrix} \\- \sum_{n=3}^\infty  \textbf{Z}_{n}\partial^n\textbf{P}\\ \Longleftrightarrow \textbf{Z}_1\nabla\vec{P}^{|z} = \Delta\vec{P} - \textbf{Z}_2\partial^2\textbf{P} - \sum_{n=3}^\infty \textbf{Z}_n\partial^n\textbf{P}
\end{multline}
\begin{equation}
    \implies \nabla\vec{P}^{|z} = (\textbf{Z}_1^T\textbf{Z}_1)^{-1}\textbf{Z}_1^T\Delta \vec{P} - (\textbf{Z}_1^T\textbf{Z}_1)^{-1}\textbf{Z}_1^{T}\textbf{Z}_2\partial^2\textbf{P} + \text{higher order terms},
\end{equation}
where $\vec{\mathcal{Z}}^{(j)} = \vec{x}^{|z_j} - \vec{x}^{|z}$, $j\in [1,M]$; $  (\Delta \vec{P})_j = \vec{P}^{|z_j}-\vec{P}^{|z}$; $ \ \textbf{Z}_n \in \mathbb{R}^{M\times (\text{dim}\vec{P})^n}$ contains $n$th order spatial distances between centroid $z$ and neighbor $z_j$; and $\partial^n\textbf{P} \in \mathbb{R}^{(\text{dim}\vec{P})^n\times  \text{dim}\vec{P}}$ contains the ordered sequence of $n$th order derivatives on $\vec{P}^{|z}$. If Eq.~\eqref{eq-gradPapprox} is linearized to $\textbf{Z}_1\nabla\vec{P}^{|z}_{lin} =$ 

\begin{figure}[ht]
    \centering
    \includegraphics[width=\linewidth]{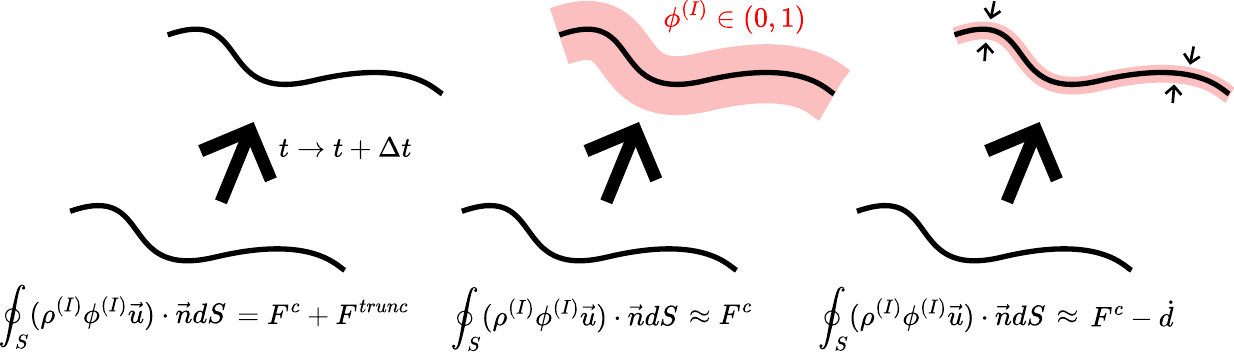}
    \caption{Diffusive impact of truncation error associated with FVM approximation and mitigation through interface sharpening.}
    \label{fig-theory-ddot}
\end{figure}

$\Delta \vec{P}$, then this introduces a componentwise truncation error of 
\begin{equation}\label{eq-type2error}
    \boldsymbol{\mathcal{E}}_{\nabla\vec{P}^{|z}} := \nabla\vec{P}^{|z} - \nabla\vec{P}^{|z}_{lin} = \mathcal{O}(\Bigl[\textbf{Z}_1^T\textbf{Z}_1\Bigl]^{-1}\textbf{Z}_1^{T}\textbf{Z}_2\partial^2\textbf{P}).
\end{equation}
Substituting Eqs. \eqref{eq-type1error},\eqref{eq-type2error} into Eq.~\eqref{eq-surfaceintapprox},
\begin{equation}\label{eq-surfaceintapprox2}
     \oint_S \vec{P}\cdot\vec{n}dS =   \underbrace{\sum_{f\in V}\vec{P}^{|f}_{lin} \cdot \vec{n}_f}_{F^{c}} + \underbrace{\Biggl[\sum_{f\in V}(\vec{\mathcal{E}}_{\vec{P}^{|f}} + \boldsymbol{\mathcal{E}}_{\nabla\vec{P}^{|z}}\cdot\vec{\mathcal{X}})\cdot\vec{n}_f\Biggl]}_{F^{trunc}},
\end{equation}
where the bracketed term is the total truncation error resulting from linearized approximation of the partial density at the faces. In this case, phase loss on the order of $F^{trunc}$ is expected and the interface is obfuscated. Therefore, it is important to develop an interface sharpening term, $\dot{d} = \mathcal{O}(-F^{trunc})$, to promote a resolved interface in Eq. ~\eqref{eq-euler2}, as shown in Fig. \ref{fig-theory-ddot}.

The left hand side of Eq.~\eqref{eq-euler2}, i.e. the choice of conserved variables, are taken from Sementilli \cite{sementilli2023scalable}. In this study, interface sharpening and surface tension are added as body forces to the right hand side of this preexisting flow formulation. For interface sharpening, start from the sharpened advection of $\phi$, i.e. $D\phi/Dt = \Gamma f_{sharp}$, where $\Gamma$ is a scaling factor and $f_{sharp}$ is the negative diffusion term
\begin{equation}\label{eq-fsharp1}
    f_{sharp}=f_{sharp}(\phi,\nabla\phi )=\nabla\cdot    \Bigl[  \varepsilon\nabla\phi - \phi(1-\phi)\frac{\nabla\phi}{||\nabla\phi||}   \Bigl]
\end{equation}
as derived in \cite{chiu2011conservative}. Then the source term for the advection of $\rho^{(I)}\phi^{(I)}\Leftrightarrow \rho^{(I)}\phi$ is approximated via
\begin{multline}
\rho^{(I)}\frac{D\phi}{Dt}=\rho^{(I)}\Gamma f_{sharp}\\ \implies \frac{\partial (\rho^{(I)}\phi )}{\partial t} + \nabla\cdot(\rho^{(I)}\vec{u}\phi) = \rho^{(I)}\Gamma f_{sharp} + \phi \frac{\partial \rho^{(I)}}{\partial t} + \phi \vec{u}\cdot\nabla\rho^{(I)}.
\end{multline}

If phase $I$ is represented by an stiffened gas equation of state, as is the case in this study, then the approximation $\partial \rho^{(I)}/\partial t \approx 0$ and $||\nabla\rho^{(I)}|| \approx 0$ holds locally to an individual cell center, leading to the source term
\begin{equation}
    \frac{\partial (\rho^{(I)}\phi)}{\partial t} + \nabla\cdot (\rho^{(I)}\phi\vec{u})=\dot{d}=\rho^{(I)}\Gamma f_{sharp}.
\end{equation}

\begin{figure}[ht]
    \centering
    \includegraphics[width=\linewidth]{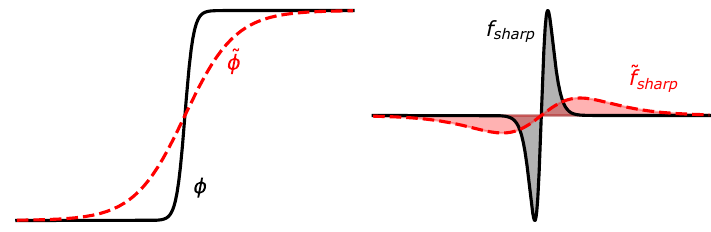}
    \caption{Convolved volume fraction field $\tilde{\phi}$ and corresponding negative diffusion term $\tilde{f}_{sharp}$.}
    \label{fig-fsharp-conserve}
\end{figure}

Notice in Eq. \ref{eq-fsharp1} that $f_{sharp}$ is a function of the volume fraction gradient, $\nabla\phi$. For sharply discontinuous $\phi$ fields, derivative computation in the direction normal to the interface creates instabilities \cite{gomez2000local}. The solution to this is to convolve the $\phi$ field with a Gaussian filter $G$, such that \begin{equation}
    \tilde{\phi}=\phi*G
\end{equation} is a smoothed auxiliary field whereby the computation of derivatives $\nabla\tilde{\phi}$ are more stable. For all initial volume fraction fields $\phi(\vec{x})$, this convolution obeys the global conservation condition
\begin{equation}
    \int_{\mathbb{R}^{\dim \vec{x}}} \tilde{f}_{sharp}(\tilde{\phi}(\vec{x}),\nabla\tilde\phi(\vec{x}))d\vec{x} = \int_{\mathbb{R}^{\dim \vec{x}}} f_{sharp}(\phi(\vec{x}), \nabla\phi(\vec{x}))d\vec{x}
\end{equation}
in which the convolved negative diffusion term is
\begin{equation}\label{eq-tildefsharp}
    \tilde{f}_{sharp}=\nabla\cdot    \Bigl[  \varepsilon\nabla\tilde{\phi} - \tilde{\phi}(1-\tilde{\phi})\frac{\nabla\tilde{\phi}}{||\nabla\tilde{\phi}||}   \Bigl]
\end{equation} as illustrated in Fig. \ref{fig-fsharp-conserve}. 

A discussion of the relationship between ($i$) choice of  Gaussian standard deviation length and ($ii$) finite choice of convolution integral boundaries on the solution accuracy is in \ref{sec-convolve}. Then the source term in its convolved form is
\begin{equation}\label{eq-ddot}
    \dot{d}= \rho^{(I)}\Gamma\tilde{f}_{sharp}.
\end{equation}

As for surface tension, the Brackbill representation of a surface volume force \cite{brackbill1992continuum}, is
\begin{equation}\label{eq-fsv1}
    \vec{F}^{sv} = \sigma \tilde{\kappa}\nabla\tilde{\phi}
\end{equation}
in which $\sigma$ is a scaling coefficient and $\tilde{\kappa}=\nabla\cdot\left(\nabla\tilde{\phi}/||  \nabla\tilde{\phi}  ||\right)$ is a modified curvature. The term in Eq.~\eqref{eq-fsv1} is added as a source term on the right hand side of the momentum density equation, and the corresponding work rate is added to the right hand side of the energy density equation. Altogether, the source vector in Eq.~\eqref{eq-euler2} is solved as
\begin{equation}\label{eq-vectorS}\vec{S}= \begin{Bmatrix}
        \dot{d}\\0\\ \rho \vec{f} + \vec{F}^{sv}\\ (\rho \vec{f} + \vec{F}^{sv})\cdot \vec{u} + \dot{q}
    \end{Bmatrix},
\end{equation} 
where $\rho \vec{f}$ indicate additional body-force terms such as gravity and $\dot{q}$ indicates a heat source, where applicable.

\subsection{Source vector discretization}\label{sec-gradcomp}

In this study, field variables stored at the centers of the finite volumes in the domain. For a structured quadrilateral grid, a standard discretization of $\tilde{f}_{sharp},$  $\vec{F}^{sv}$ in source vector $\vec{S}$ (Eq.~\eqref{eq-vectorS}) would include notions of $\Delta x_i$, predicated on the assumptions that $(i)$ cell centers are equally spaced, such that $\Delta x_i$ are constant; and $(ii)$ cell edges are mutually perpendicular, such that normal their vectors align with coordinate axes. However both of these assumptions break when not using structured quadrilaterals. Therefore the purpose of this section is to develop methods for discretizing $\nabla\tilde{\phi} \rightarrow \tilde{f}_{sharp}(\nabla\tilde{\phi}), \ \vec{F}^{sv}(\nabla\tilde{\phi})$  that are generalizable to an arbitrary cell center distribution.

\subsubsection{Gradient computation}

\begin{figure}[ht]
    \centering
    \begin{subfigure}[c]{0.35\linewidth}
        \includegraphics[width=\linewidth]{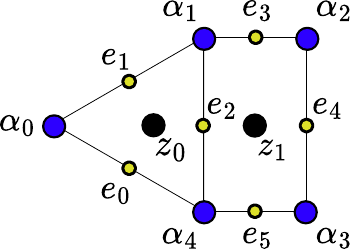}    
        \caption{\ }
        \label{fig-theory-mesh}
    \end{subfigure}\hspace{2cm}
    \begin{subfigure}[c]{0.35\linewidth}
        \includegraphics[width=\linewidth]{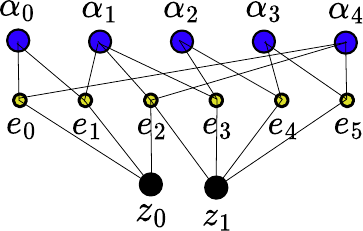}    
        \caption{\ }
        \label{fig-dag1}
    \end{subfigure}
    \caption{Isomorphism between an arbitrarily constructed mesh and its Hasse diagram/DAG. Centers are $z_j$ (black), edges are $e_j$ (yellow), and vertices are $\alpha_j$ (blue). (a): Mesh. (b): DAG.}
    \label{fig-theory-dag}
\end{figure}

Given that $f_{sharp}=f_{sharp}(\tilde{\phi},\nabla\tilde{\phi})$, it is necessary to be able to compute derivatives at points. For a particular point, derivative calculation requires knowledge of what constitutes a neighbor, which for non-quadrilateral grids is no longer defined by distance. However, if it is known whether each point in the domain is a center, edge, or vertex, then a topological approach can be used to deduce the structure of each cell. In particular points can be mapped in terms of a Hasse diagram/directed acrylic graph (DAG) which establishes transitive covering relationships of the cells. Since there is an isomorphism between any mesh (Fig. \ref{fig-theory-mesh}) and its representative DAG (Fig. \ref{fig-theory-dag}), the covering relationships are sufficient to define the mesh's cell structure. This study uses PETSc DMPlex's mesh generation and isomorphic DAG transformation capabilities \cite{petsc-efficient,petsc-user-ref,petsc-web-page}.

\begin{figure}[ht]
    \centering
     
        \begin{subfigure}[c]{0.3\linewidth}
            \includegraphics[width=\linewidth]{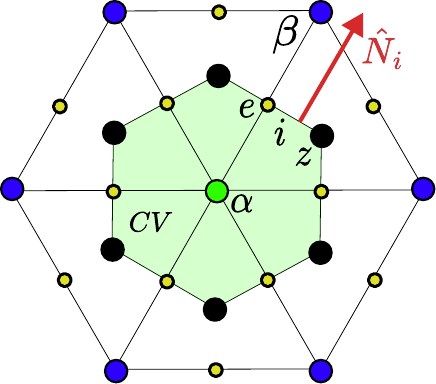}    
            \caption{\ }
            \label{fig-gvfv}
        \end{subfigure}
        \hspace{1cm} 
        \begin{subfigure}[c]{0.3\linewidth}
            \includegraphics[width=\linewidth]{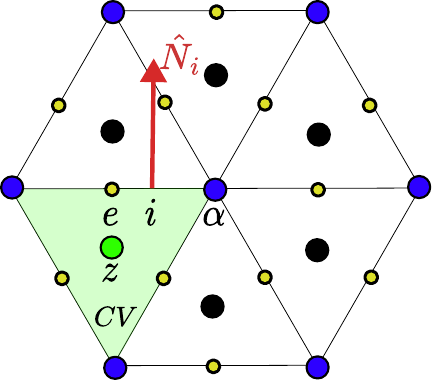}    
            \caption{\ }
            \label{fig-gcfv}
        \end{subfigure}
        \\
        \bigskip
        \begin{subfigure}[c]{0.3\linewidth}
            \includegraphics[width=\linewidth]{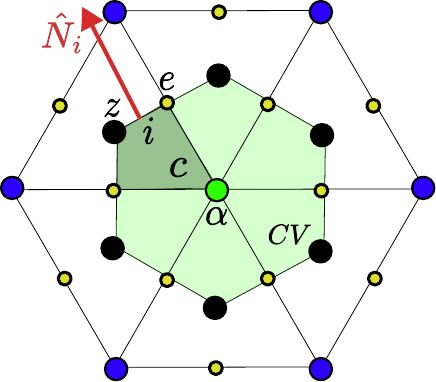}    
            \caption{\ }
            \label{fig-gvfc}
        \end{subfigure}
        \hspace{1cm}
        \begin{subfigure}[c]{0.3\linewidth}
            \includegraphics[width=\linewidth]{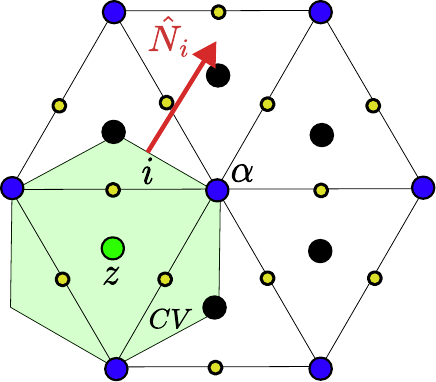}    
            \caption{\ }
            \label{fig-gcfc}
        \end{subfigure}
    
    \caption{Method of obtaining the gradient of a field variable evaluated at \rule{1cm}{0.1mm} sourced from field values stored at \rule{1cm}{0.1mm}. Centers are $z$, edges are $e$, vertices are $\alpha,\beta$, facets are $i$, facet normals are $\hat{N}_i$, and corners are $c$. Note, corners are bounded by a vertex, a center, and the connecting edges. (a): \textbf{vertex}, \textbf{vertices}; (b): \textbf{center}, \textbf{vertices}; (c): \textbf{vertex}, \textbf{centers}; (d): \textbf{center}, \textbf{centers}.}
    \label{fig-gxfx}
\end{figure}
From the DAG it is possible to deduce each point's cloud of cell neighbors or cloud of vertex neighbor. Then the gradient of an arbitrary scalar field $b=b(\vec{x})$ evaluated at that point is 
\begin{multline}
    \nabla b\Bigl|_{point} = \frac{1}{V_{CV}}\sum_{neighbors\in CV}\hat{N}_{neighbor} \ b_{neighbor}, \\  \hat{N}_{neighbor} = \sum_{facets\in neighbor} \hat{N}_{facet}.
\end{multline}
Here a facet is a line segment connecting one of the point's neighbors to an adjacent edge, $V_{CV}$ represents the control volume bounded by the set of facets, and  $\hat{N}_{facet}$ represents the outward normal of the facet \cite{morgan20173d}. The formation of the control volume and the facets depends on $(i)$ whether the point $\mathscr{P}_1$ is a center or a vertex, and $(ii)$ whether the cloud of neighbors $\mathscr{P}_2$ is center-based or vertex-based. Each permutation is visualized in Fig. \ref{fig-gxfx}, and each corresponding gradient computation is written in Table \ref{tab-getgrad}, in which $\{\mathscr{P}_1\in (\text{vertex, center}),\mathscr{P}_2\in (\text{vertices, centers})\}$.

\begin{table}[ht]
\centering
\renewcommand{\arraystretch}{1.2}
\noindent\makebox[\linewidth]{
\resizebox{1\textwidth}{!}{
\begin{tabular}{|p{6cm}|c|p{7.5cm}|}
\hline
$\mathscr{P}_1, \mathscr{P}_2 :\nabla b\Bigl|_{\mathscr{P}_1}=f[b(\sum \mathscr{P}_2)]$ & Fig. & Formula \\ \hline
$\mathscr{P}_1=$ vertex, $\mathscr{P}_2=$ vertices & 4(a) & $\begin{array}{l}  
\nabla b\Bigl|_{\alpha} = \frac{1}{V_{CV}} \sum_{e\in CV} \hat{N}_e b_e \\  
\hat{N}_e = \sum_{i\in e}\hat{N}_i \\  
b_e=(b_\alpha+b_\beta)/2  
\end{array}$ \\ \hline
$\mathscr{P}_1=$ center, $\mathscr{P}_2=$ vertices & 4(b) & $\begin{array}{l}  
\nabla b\Bigl|_{z} = \frac{1}{V_{CV}}\sum_{\alpha \in CV} \hat{N}_\alpha b_{\alpha} \\  
\hat{N}_\alpha = \sum_{i\in \alpha}\hat{N}_i  
\end{array}$ \\ \hline
$\mathscr{P}_1=$ vertex, $\mathscr{P}_2=$ centers & 4(c) & $\begin{array}{l}  
\nabla b\Bigl|_{\alpha} = \frac{1}{V_{CV}} \sum_{c\in CV} \hat{N}_c b_{z(c)} \\  
\hat{N}_c = \sum_{i\in c}\hat{N}_i  
\end{array}$ \\ \hline
$\mathscr{P}_1=$ center, $\mathscr{P}_2=$ centers & 4(d) & $\begin{array}{l}  
\nabla b\Bigl|_{z} = \frac{1}{V_{CV}} \sum_{z\in CV}\hat{N}_z b_{z} \\  
\hat{N}_z = \sum_{i\in z}\hat{N}_i  
\end{array}$ \\ \hline
\end{tabular}
}}
\caption{}
\label{tab-getgrad}
\end{table}
\FloatBarrier
Using the strategy described in Table \ref{tab-getgrad} the calculation of source terms is described in the rest of the section.

\subsubsection{Calculation of negative diffusion field}

Let the convolved volume fraction field $\tilde{\phi}$ be stored at cell centers, $z$, and let vertices be denoted as $\alpha$. Then the process of obtaining $(\tilde{f}_{sharp})_z$ in Eq.~\eqref{eq-fsharp1} at centers is two-fold. $(i)$ First, $\nabla\tilde{\phi}_\alpha$ is computed at vertices based on the cell-centered smoothed volume fraction information, $\tilde{\phi}_z$, using the method shown in Fig. \ref{fig-gvfc}. Then
\begin{equation}
    \vec{a}_\alpha =  \Bigl[  \varepsilon\nabla\tilde{\phi}_\alpha - \tilde{\phi}_\alpha(1-\tilde{\phi}_{\alpha})\frac{\nabla\tilde{\phi}_{\alpha}}{||\nabla\tilde{\phi}_{\alpha}||}    \Bigl] = \sum_j a^{(j)}_\alpha \hat{e}_j,
\end{equation}
in which $\phi_\alpha$ is the weighted average of neighboring $\phi_z$ and $\hat{e}_j$ denotes basis vectors. $(ii)$ Second, $(\tilde{f}_{sharp})_z=(\nabla\cdot \vec{a})_z$ is computed at centers based on the diagonal components of the gradients of the vertex-based scalar fields, $a_\alpha^{(j)}\in \vec{a}_\alpha$, using the method shown in Fig. \ref{fig-gcfv}. That is,
\begin{equation}
    (f_{sharp})_z = \Tr\begin{Bmatrix}
        \nabla (a^{(1)}_\alpha) & \nabla (a^{(2)}_\alpha)  & \nabla (a^{(3)}_\alpha)
    \end{Bmatrix} \implies \dot{d}_z = \Gamma_1\rho^{(I)}(\tilde{f}_{sharp})_z.
\end{equation}

\subsubsection{Calculation of surface tension field}

Obtaining the surface force field requires computing the cell-centered curvature and volume fraction gradient, as in Eq.~\eqref{eq-fsv1}. Let the smoothed volume fraction $\tilde{\phi}$ be stored at cell centers, $z$, and let vertices be denoted as $\alpha$. Then the process of obtaining curvature at centers is two-fold: ($i$) First, the vector field $\vec{n}_\alpha$ is computed at vertices based on the cell-centered smooth volume fraction information, $\tilde{\phi}$, using the method shown in Fig. \ref{fig-gvfc}. ($ii$) Second, $\tilde{\kappa}_z$ is computed at centers based on the diagonal components of the gradients of the vertex-based scalar fields, $n_\alpha^{(j)}\in \vec{n}_\alpha$, using the method shown in Fig. \ref{fig-gcfv}. That is, 
\begin{equation}
    \tilde{\kappa}_z = \Tr\begin{Bmatrix}
        \nabla (n^{(1)}_\alpha) & \nabla (n^{(2)}_\alpha)  & \nabla (n^{(3)}_\alpha)
    \end{Bmatrix}.
\end{equation}
Finally, $\nabla\tilde{\phi}_z$ is computed at centers based on cell-centered smooth volume fraction information, $\tilde{\phi}$, using the method shown in Fig. \ref{fig-gcfc}.
\subsection{System closure}\label{sec-closure}

The compressible flow system presented by Eq.~\eqref{eq-euler2} tracks the partial density of one of the phases, as well as the mixture mass density, mixture momentum density, and mixture energy density ($\rho^{(I)}\phi, \rho, \rho \vec{u}, \rho e$) per unit volume. However, this system does not inform the  pressure, temperature, volume fraction, and density and internal energy of each individual phase ($p,T,\phi,\rho^{(I),(II)},e^{(I),(II)}$). Auxiliary variables such as these are resolved by establishing thermodynamic equations of state that each phase obeys (assuming that the  molecules have negligible volume and that molecule collisions are perfectly elastic) \cite{jibben2019modeling}. As shown in Eq.~\eqref{eq-eos}, each of the fluids are modeled by the stiffened equations of state\begin{equation}\label{eq-eos}
    \begin{cases}
        p=(\gamma-1)\rho e - \gamma p^0,\\
        a=\sqrt{\gamma(p+p^0)}/\rho, \\
        e=C_pT/\gamma + p^0/\rho. 
    \end{cases}
\end{equation}
In addition the notion of mass fraction can be taken directly from the conserved variables as $Y^{(I)}=\phi\rho^{(I)}/\rho , \ \ Y^{(II)}=1-Y^{(I)}.$
Mass fraction weighs the relative contribution of each phase towards the total mixture density and total mixture internal energy. The equations representing these weights are
\begin{equation}\label{eq-1rhoygrhogylrhol}
    \frac{1}{\rho} = \frac{Y^{(I)}}{\rho^{(I)}} + \frac{Y^{(II)}}{\rho^{(II)}},
\end{equation}
\begin{equation}\label{eq-eygegylel}
    e = Y^{(I)}e^{(I)} + Y^{(II)}e^{(II)}.
\end{equation}
Then, the auxiliary variables $\vec{A}=\{\rho^{(I)},\rho^{(II)},T:=T^{(I)}=T^{(II)}\}$ are obtained by minimizing the residual of the system of equations formed by $p^{(I)}=p^{(II)}$ and Eqs.~\eqref{eq-1rhoygrhogylrhol}, ~\eqref{eq-eygegylel}. The residual is $\vec{r}(\vec{A})=\{f_1(\vec{A}), f_2(\vec{A}), \ f_3(\vec{A}) \ \}^T\rightarrow \vec{0}$, where
\begin{multline*}
    f_1(\vec{A})=\gamma^{(II)}(\gamma^{(I)}-1)(\rho C_{p}^{(I)}T + \gamma^{(I)} p^{0(I)})\rho^{(I)} - \rho \gamma^{(I)2} \gamma^{(II)} p^{0(I)} \\ - \gamma^{(I)} (\gamma^{(II)} - 1)(\rho C_{p}^{(II)} T + \gamma^{(II)} p^{0(II)})\rho^{(II)} + \gamma^{(II)2} \gamma^{(I)} p^{0(II)} \rho,
\end{multline*}
\begin{equation*}
    f^{(II)}(\vec{A})= Y^{(I)} \rho \rho^{(II)} + Y^{(II)} \rho \rho^{(I)} - \rho^{(I)} \rho^{(II)},
\end{equation*}
\begin{multline}
    f_3(\vec{A})= Y^{(I)} \gamma^{(II)} (\rho C_{p}^{(I)}T + \gamma^{(I)} p^{0(I)}) \\+ Y^{(II)} \gamma^{(I)} (\rho C_{p}^{(II)} T + \gamma^{(II)} p^{0(II)}) - \rho \gamma^{(II)} \gamma^{(I)} e.
\end{multline}
The auxiliary variables are updated via the Newton Raphson method $\vec{A}^{k+1}=\vec{A}^k - \textbf{J}^{-1}\vec{r}(\vec{A}^k)$, $k=0,1,\dotsc $,  $J_{ij}=\partial f_i(\vec{A}^k)/\partial A^k_j$. This study uses PETSc's Newton Raphson implementation   \cite{petsc-efficient,petsc-user-ref,petsc-web-page}. 

Further, volume fraction is computed as $\phi=[\rho^{(I)}\phi]/\rho^{(I)}$, and auxiliary variables $e^{(I)},e^{(II)},p, $ can be  backed out of the equations of state according to Eq.~\eqref{eq-eos}. However, consider that the pressure EOS, $p=f(\rho)=(\gamma-1)\rho e - \gamma p^0,$ is numerically sensitive to the density solution. For example apply the thermodynamic constants of water at room temperature, $\{T,\gamma,p^{0},C_p\} = \{300\text{ K},1.932,1.1639\times10^9\text{ Pa},8095.08\text{ Jkg}^{-1}\text{K}^{-1}\}, $ to get $f(\rho=998.23 \text{ kgm}^{-3})=10^5\text{ Pa}$. If the density is changed only by $-0.01$ percent, then the pressure is computed as $f(\rho=998.10 \text{ kgm}^{-3})=-5.23\times10^4\text{ Pa}$, which is unphysical. This demonstrates a need for a pressure relaxation function to smooth out pressure solutions where the density is underestimated, such that pressure cannot be calculated as negative. Therefore in this step the pressure EOS is modified as
\begin{equation}\label{eq-pressurefix}
    p(\rho)=\begin{cases}
        \frac{\psi_1}{\psi_2-\rho}+\psi_3 & \rho\leq \bar{\rho}\\
        (\gamma-1)\rho e - \gamma p^0 & \rho > \bar{\rho}
    \end{cases}
\end{equation}
in which $\psi_1=\bar{p}'\bar{p}^2\bar{\rho}^2/(\bar{p}-\bar{p}'\bar{\rho}), \ \psi_2=-\bar{p}\bar{\rho}^2/(\bar{p}-\bar{p}'\bar{\rho}), \ \psi_3=-\bar{p}^2/(\bar{p}-\bar{p}'\bar{\rho})$, $\bar{p}=(\gamma-1)\bar{\rho}e - \gamma p^0$, $\bar{p}'=(\gamma-1)\bar{\rho}e$, and $\bar{\rho}$ is an expected density magnitude corresponding to a fluid of interest (for water, $\bar{\rho}=998.23 \text{ kgm}^{-3}$ is used). Eq.~\eqref{eq-pressurefix} satisfies the consistency conditions $p(0)=0,p(\bar{\rho})=\bar{p},$ and $dp(\bar{\rho})/d\rho = \bar{p}'$.

\subsection{Flux evaluation}\label{sec-fluxeval}
The fluxes in the compressible multiphase flow system used in this study are written in strong form in the second term of the left hand side in Eq.~\eqref{eq-euler2}. Integrate with respect to a cell volume $V$ and apply the divergence theorem to get
\begin{equation}\label{eq-fluxcalc1}
    \int_V  \nabla\cdot \begin{Bmatrix}
        \rho^{(I)}\phi\vec{u}\\\rho\vec{u}\\\rho\vec{u}\otimes\vec{u} + p\textbf{I} - \boldsymbol{\tau}\\\rho e \vec{u} - \boldsymbol{\tau}\cdot\vec{u}
    \end{Bmatrix}dV = \oint_S \begin{Bmatrix}
        \rho^{(I)}\phi(\vec{u}\cdot\vec{n})\\
        \rho (\vec{u}\cdot\vec{n})\\
        \rho (\vec{u}\cdot\vec{n}) \vec{u} + p\vec{n} - \boldsymbol{\tau}\cdot\vec{n}\\
        \rho (\vec{u}\cdot\vec{n})H-(\boldsymbol{\tau}\cdot\vec{u})\cdot\vec{n}
    \end{Bmatrix}dS
\end{equation}
where $S$ is the cell volume's surface and $\vec{n}$ represents the surface normal. In finite volume methods, surface integrals are discretized as summations of field variables along cell volume faces, in the manner of Eq.~\eqref{eq-surfaceintapprox}. However, field variables are stored in cell centers, meaning that some method of evaluating all notions of mass flux ($\dot{m}:=\rho(\vec{u}\cdot\vec{n})$) and pressure ($p$) at the faces (denoted by subscript $_{1/2}$) must be introduced. A popular solution to this is an advection upstream splitting method with additional pressure and velocity diffusion terms (AUSM+up), which originates in work by Liou et al \cite{liou1993new,liou1996sequel,liou2006sequel}. This method requires the mixture speed of sound on the left and right sides (denoted by subscripts $_{L/R}$) as input. In this study they are approximated via
\begin{equation}
    a_{L/R} = \sqrt{\frac{1}{\phi_{L/R}/a_{L/R}^{(I)2} + (1-\phi_{L/R})/a_{L/R}^{(II)2}} }.
\end{equation}
It follows that the left and right mixture Mach numbers are $M_{L/R}=u_{L/R}/a_{1/2}$, where $u_{L/R}$ is the velocity component normal to the left and right faces, and $a_{1/2}=(a_L+a_R)/2$. The mean mixture Mach number across the interface is approximated as 
\begin{equation}
    M_o=\sqrt{\min[1,\max\left(\overline{M}^2,M_\infty^2\right)},
\end{equation}
where
\begin{equation*}
    \overline{M}^2=\frac{u_L^2+u_R^2}{2a_{1/2}}
\end{equation*}
and $M_\infty$ is a constant representing the expected Mach number magnitude. Then the interfacial mixture Mach number is
\begin{equation}
    M_{1/2} = f_4^+(M_L) + f_4^-(M_R)\\-\frac{K_p}{f_a}\max\left( 1-\sigma_M\overline{M}^2,0  \right)\frac{p_R-p_L}{\rho_{1/2}a^2_{1/2}},
\end{equation}
in which
\begin{equation*}
    f^\pm_4(x) = \begin{cases}
        f^\pm_1(x) & |x|\geq 1,\\
        [1\mp 16\beta f_2^\mp (x)]f^\pm_2(x) & |x|<1,
    \end{cases}
\end{equation*}
\begin{equation*}
    f_1^\pm(x)=\frac{1}{2}(x\pm|x|),
\end{equation*}
\begin{equation*}
    f_2^\pm(x) = \pm\frac{1}{4}(x\pm1)^2,
\end{equation*}
\begin{equation*}
    f_a=M_o(2-M_o),
\end{equation*}
constants $\{K_p,\sigma_M,\beta\}=\{0.25,1.0,0.125\}$, and $\rho_{1/2}=(\rho_L+\rho_R)/2$. The interfacial mass flux is calculated as 
\begin{equation}\label{eq-massfluxface}
    \dot{m}_{1/2}=a_{1/2}M_{1/2}\begin{cases}
        \rho_L & M_{1/2}>0,\\
        \rho_R & M_{1/2}\leq 0,
    \end{cases}
\end{equation}
and the interfacial pressure is
\begin{equation}\label{eq-pressureface}
    p_{1/2} = p_Lf_5^+(M_L) + p_Rf_5^-(M_R) \\ - \frac{3}{4}(\rho_L+\rho_R)(f_aa_{1/2})(u_R-u_L)f_5^+(M_L)f_5^-(M_R),
\end{equation}
in which
\begin{equation*}
    f_5^\pm(x) = \begin{cases}
        \frac{1}{x}f_1^\pm(x) & |x|\geq 1,\\
        [(\pm 2-x)\mp 3(-4+5f_a^2)xf_2^\mp (x)]  f_2^\pm (x) & |x|<1.
    \end{cases}
\end{equation*}
With Eqs.~\eqref{eq-massfluxface},~\eqref{eq-pressureface}, mass flux and pressure terms are solved at faces and can be substituted into the flux vector in Eq.~\eqref{eq-fluxcalc1} as the surface integral is discretized.
\section{Results}\label{sec-results}

This section is divided into two parts. Sec. \ref{sec-validation} demonstrates the spatial convergence and accuracy of the interface sharpening and surface tension models by testing the solver against a set of test cases. Sec. \ref{sec-droplet} applies the combination of these models to a series of Weber and Ohnesorge number-dependent fluid flows with the purpose of quantifying droplet shape and size as a function of these nondimensional parameters.

\subsection{Validation}\label{sec-validation}

\subsubsection{Curvature convergence}\label{sec-convg}

\begin{figure}[htbp]
    \centering

    \includegraphics[width=0.6\linewidth]{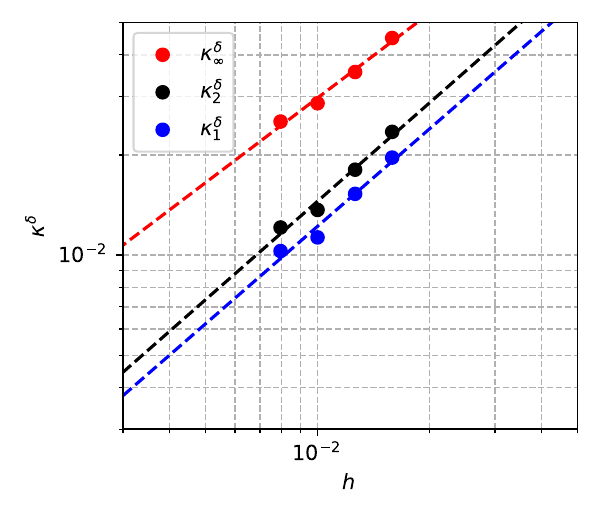}

    \caption{Convergence analysis for curvature calculation.}
    \label{fig-curvature-convg-3}
\end{figure}

To assess the convergence properties of the interface curvature calculation a grid refinement study is conducted on a unit sphere with uniform surface curvature, $\kappa=1$. With each successive test the domain is populated with increasingly refined grid spacing magnitudes $h\in [10^{-3},10^{-1}]$, with the expectation that curvature solutions become more accurate as the mesh is refined. The metric used to quantify error is

\begin{equation}\label{eq-def-kappadelta}
    \kappa^\delta_{\{1,2,\infty \}} := \Bigl|\Bigl| \kappa_{sim}\bigl|_{r=1} - \kappa_{true}\bigl|_{r=1} \Bigl|\Bigl|_{\{1,2,\infty \}},
\end{equation}
which quantifies the difference between the curvature evaluated at $r=1$ by the simulation and the analytical curvature at the same location. This metric is visualized as a function of $h$ in Fig. \ref{fig-curvature-convg-3}. The results indicate that the
$\kappa^\delta_1, \kappa^\delta_2$, $\kappa^\delta_\infty$ convergence orders are 0.97, 0.98, 0.85 respectively, indicating convergence to approximately first order.

\color{black}

\subsubsection{Static, initially diffuse circle}\label{sec-sidi}

The transformation of an initially diffuse circular interface into one that is sharp is the most straightforward test of the interface sharpening term proposed in Eq. \ref{eq-ddot}. As the negative diffusion term acts in a direction determined by the gradient of the volume fraction field, it is intuitive to pose an initial condition that gradually transitions from $\phi=1.0$ to $\phi=0.0$ in the outward radial direction. Let the subscript $_0$ indicate initial conditions. Then the diffuse interface is initialized via
\begin{equation}
    \phi_0(r) = \frac{1}{2}\biggl[  \tanh\biggl( \frac{R+r}{2\varepsilon_0}\biggl) + \tanh \biggl( \frac{R-r}{2\varepsilon_0}  \biggl)     \biggl]
\end{equation}
in which $r=\sqrt{x^2+y^2}$; $x,y\in [-0.05,0.05]$; $h=10^{-3}$; $\varepsilon_0=8h$ scales the thickness of the initial transition region; and $R=0.025$ is the radius of the $\phi=0.5$ contour. Homogeneous Neumann boundary conditions are applied for all four sides for all variables.
 
The initial conditions are $\gamma^{(I)}=1.400,\gamma^{(II)}=1.932,\rho_0^{(I)}=1.16,\rho_0^{(II)}=998.23,C_p^{(I)}=1004.50,C_p^{(II)}=8095.08,p^{0(I)}=0, p^{0(II)}=1.1645\times10^{9}$,  
and let $\Delta t, t_f$ be $10^{-7}$ and $ 500\Delta t$ respectively. This study compares results between quadrilateral grid elements and simplex grid elements, both bounded in size by the spacing $h$.

\begin{figure}[ht]
    \centering
    \hspace*{-0.9cm}
    \scalebox{0.75}{
    \begin{minipage}{\textwidth}
        \centering
        \hspace*{2.5cm}
        \begin{overpic}[width=0.5\linewidth,tics=10]{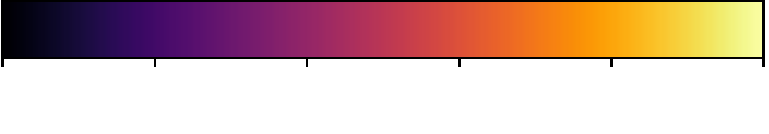}
        \end{overpic}

        \begin{overpic}[width=1.2\linewidth,tics=10]{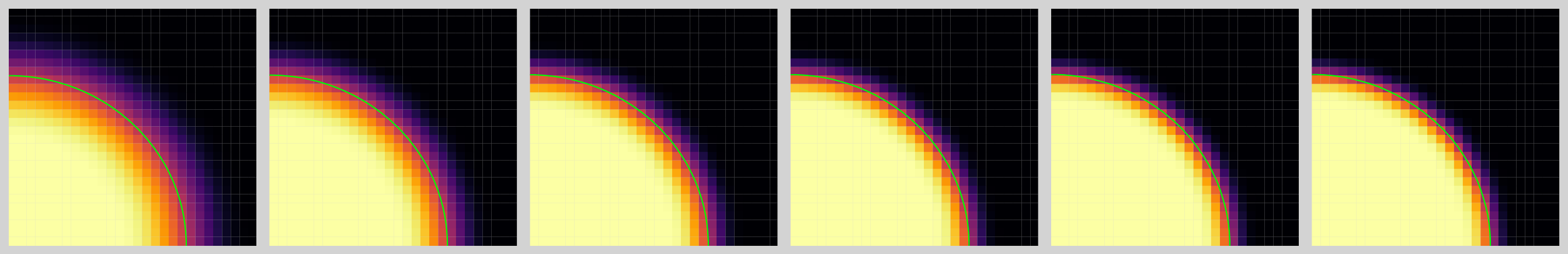}
        \end{overpic}

        \begin{overpic}[width=1.2\linewidth,tics=10]{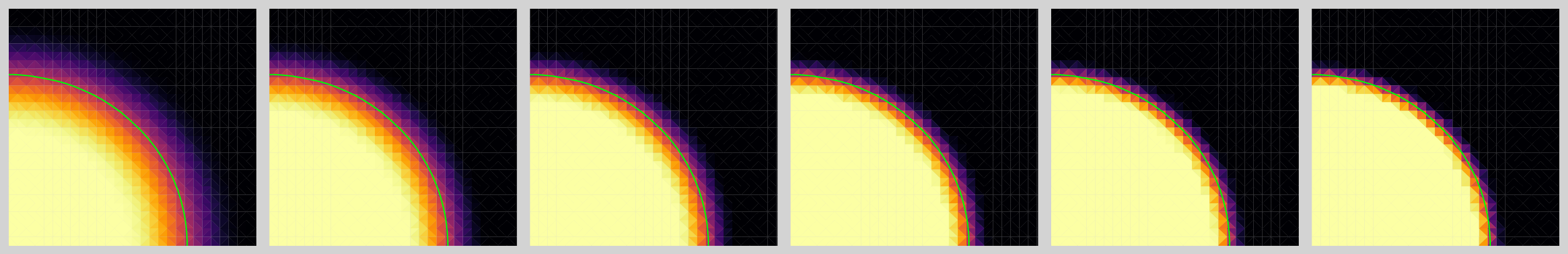}
            \put (-11, 29){\large \color{black}$t/t_f=$\color{black}}
            \put (2,29) {\large \color{white}$0.0 $\color{black}}
            \put (18,29) {\large \color{white}$0.2$\color{black}}
            \put (35,29) {\large \color{white}$0.4$\color{black}}
            \put (52,29) {\large \color{white}$0.6$\color{black}}
            \put (68,29) {\large \color{white}$0.8$\color{black}}
            \put (84,29) {\large \color{white}$1.0$\color{black}}

            \put (27.25,33) {\large $0.0$}
            \put (34.5,33) {\large $0.2$}
            \put (48.5,33.5) {\large $\phi $}
            \put (42.75,33) {\large $0.4$}
            \put (52.25,33) {\large $0.6$}
            \put (61.5,33) {\large $0.8$}
            \put (68.25,33) {\large $1.0$}

            \put (-11, 13){\large \color{black}$t/t_f=$\color{black}}
            \put (2,13) {\large \color{white}$0.0$\color{black}}
            \put (18,13) {\large \color{white}$0.2$\color{black}}
            \put (35,13) {\large \color{white}$0.4$\color{black}}
            \put (52,13) {\large \color{white}$0.6$\color{black}}
            \put (68,13) {\large \color{white}$0.8$\color{black}}
            \put (84,13) {\large \color{white}$1.0$\color{black}}

        \end{overpic}

        \caption{\large{Time-evolved $\phi$ field of a diffuse circular body subjected to interface sharpening only. Green lines indicate position of final $\phi=0.5$ contour. Top: quadrilateral mesh. Bottom: simplex mesh.}}
        \label{fig-sidc}
    \end{minipage}
    }
\end{figure}

\begin{figure}[ht]
    \centering
    \makebox[\textwidth]{
\includegraphics[width=\linewidth]{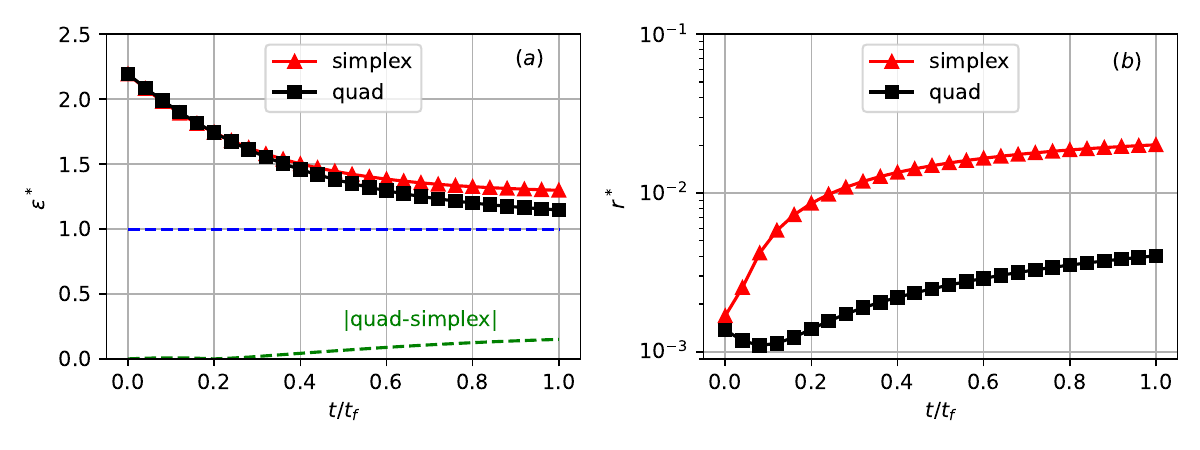}
    }
    \caption{$(a)$ Time-evolved normalized interface thickness parameter with asymptote $\varepsilon^*\longrightarrow 1$, indicating the convergence of interfacial thickness to a single cut cell. Blue line indicates asymptote and green line indicates difference in thickness between quad and simplex meshes.  $(b)$ Time-evolved normalized radial standard deviation $r^*\ll 1$ indicating low interface distortion.}
    \label{fig-quadsimplex}
\end{figure}

Fig. \ref{fig-sidc} displays the time-evolved volume fraction field in which the preimposed transition region converges to a binary jump, both on a mesh of quadrilaterals and on simplices. To quantify both convergence rates, the two metrics  \begin{equation}\label{eq-def-rstar}
    \varepsilon^* := \frac{\Bigl|\max (r\bigl|_{\phi=0.5\pm 0.25})  - \min (r\bigl|_{\phi = 0.5\mp 0.25}) \Bigl|}{h}, \ \ \ \ r^* := \frac{\text{std}\Bigl(r\Bigl|_{\phi=0.5}\Bigl)}{h}
\end{equation}
are introduced as measures of transition region length and radius shape respectively. In Eq.~\eqref{eq-def-rstar}, notions of $r\bigl|_{\phi\in [0,1]}$ refer to the radii of contours corresponding to a specific $\phi$ value in the domain. The contours are obtained via Matplotlib interpolation capabilities. Both metrics $\varepsilon^*,r^*$ in Eq.~\eqref{eq-def-rstar} are normalized by the grid spacing magnitude, $h$, to inform the number of grid spaces across which $(i)$ the interface stretches radially outward, in the case of $\varepsilon^*$, and $(ii)$ the radius deviates along the surface, in the case of $r^*$. The first expected behavior in time is $\varepsilon^*\longrightarrow 1$, representing the convergence to an interface subsumed by a single cut cell. This is seen in Fig. \ref{fig-quadsimplex}$(a)$ for both the quadrilateral mesh and the simplex mesh case. The second expected behavior in time is $r^*\approx  $ constant, representing shape retention in the absence of a velocity field. In Fig. \ref{fig-quadsimplex}$(b)$ the case corresponding to the simplex mesh radially deviates at one order of magnitude higher than the quadrilateral mesh. However the deviation is bounded by $\text{std}(r|_{\phi=0.5})\leq \mathcal{O}(  10^{-2}h) \ll h$ for all $t\in [0,t_f]$.

\FloatBarrier

\subsubsection{Zalesak disk}\label{sec-zalesak}

\begin{figure}[ht]
    \centering
    
    \begin{subfigure}[t]{0.49\textwidth}
        \centering
        \includegraphics[width=\linewidth]{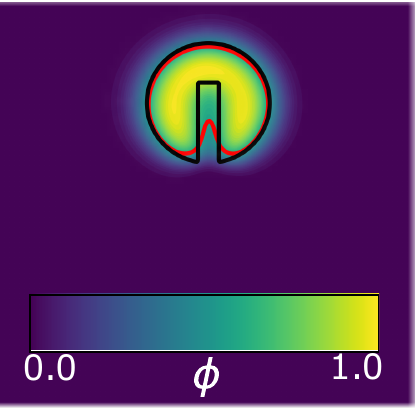}
        \caption{}
        \label{fig:zalesak-nois}
    \end{subfigure}
    \hfill
    \begin{subfigure}[t]{0.49\textwidth}
        \centering
        \includegraphics[width=\linewidth]{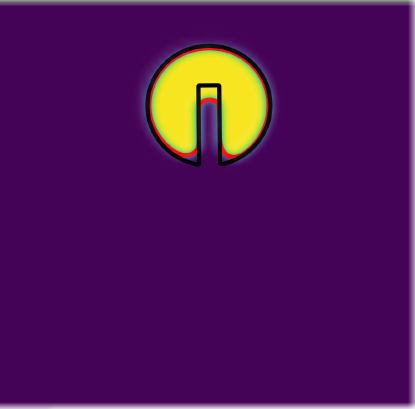}
        \caption{}
        \label{fig:zalesak-is}
    \end{subfigure}
    
    \caption{Simulated Zalesak disk after one revolution for cases (a) without and (b) with interface sharpening.}
    \label{fig:zalesak-overview}
\end{figure}

\begin{figure}[htbp]
    \centering
    \begin{subfigure}[t]{\linewidth}
        \centering
        \includegraphics[width=\textwidth]{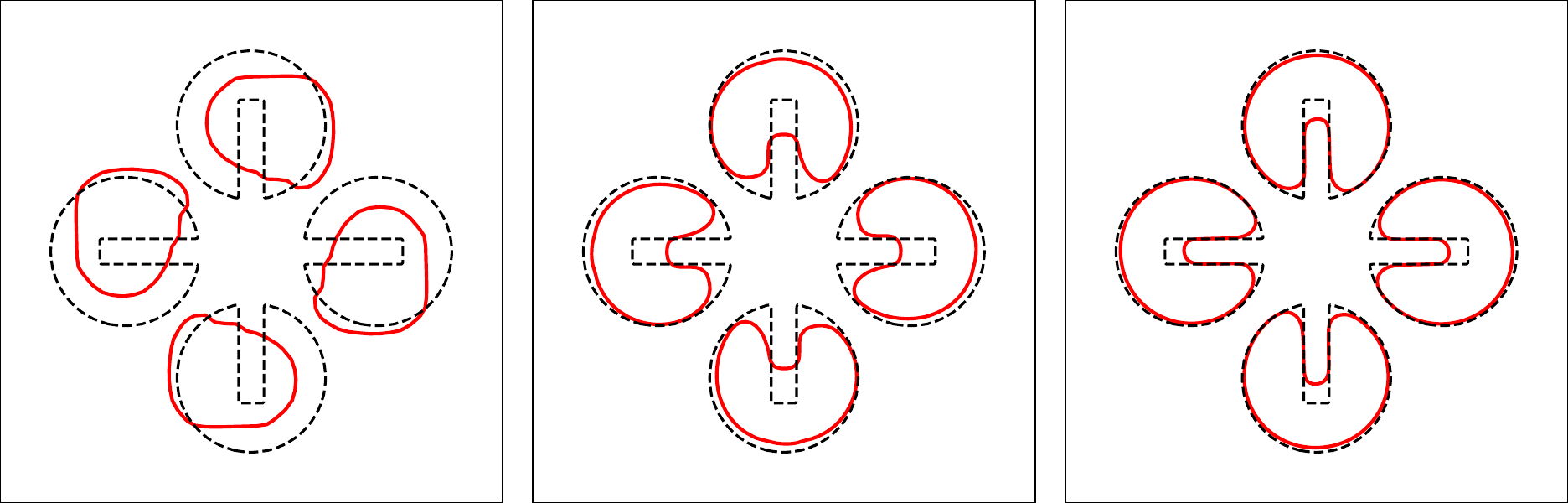}
        \caption{}
        \label{fig:zalesak-contours}
    \end{subfigure}
    \vspace{1.2em}
    \begin{subfigure}[t]{\textwidth}
        \centering
        \includegraphics[width=0.8\textwidth]{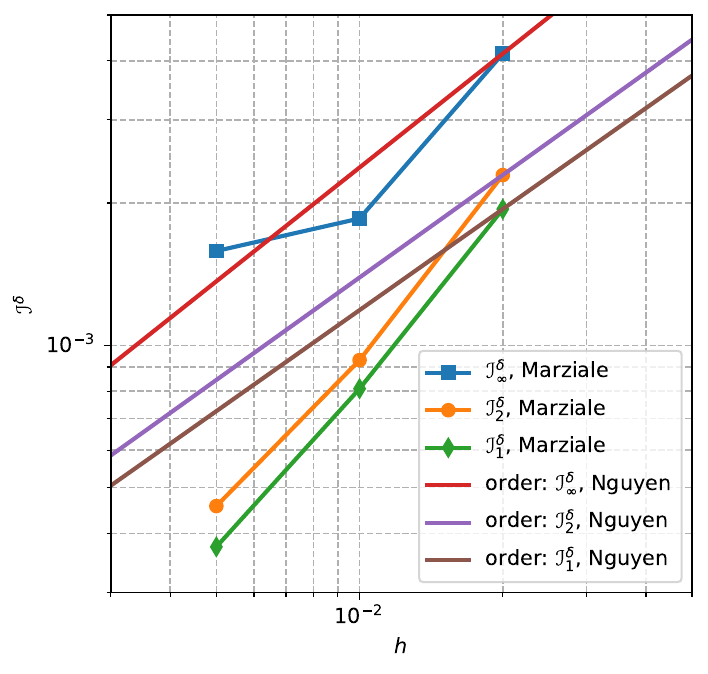}
        \caption{}
        \label{fig:zalesak-convergence}
    \end{subfigure}
    \caption{
    $(a)$ Interfacial evolution of Zalesak disk for grid spacings $h\sim 2\times 10^{-2}, 10^{-2}, 5\times 10^{-3}$ (from left to right), compared to the ideal case. $(b)$ Convergence order of interfacial error, quantified using Eq.~\eqref{eq-def-Idelta} and compared to slopes computed from results reported by Nguyen \cite{nguyen2017volume}. }
    \label{fig:zalesak-quantitative}
\end{figure}

The Zalesak disc is a classic test of the fidelity of interface sharpening schemes featuring advection \cite{nguyen2017volume,zalesak1979fully,zu2020phase,shen2024enhanced,vijay2025comparative}. The conditions of the test are as follows: Initialize a slotted circular disk characterized by phase $I$, embedded in the domain $x,y\in [0,1]$. Center the disk at $(0.5,0.75)$ with radius $0.15$. Insert a rectangular slot characterized by the background phase, i.e. phase $II$, with width $0.05$ and depth $0.25$. Impose the constant velocity field $\vec{u}(t) =  2\pi \{ (-y+0.5), (x-0.5)\}^T/t_f$ to complete one full revolution over the time interval $t\in [0,t_f]$. The ideal solution returns the interface of the slotted disk to its original position and shape (i.e. at $t=0$), with no loss of sharpness of the slot. The disk's simulated behavior can be judged against this ideal solution according to the 1-, 2-, and $\infty$-norms
\begin{equation}\label{eq-def-Idelta}
    \mathscr{I}^\delta _{\{1,2,\infty \}} := \Bigl|\Bigl| \mathscr{I}_{sim}-\mathscr{I}_{true} \Bigl|\Bigl|_{\{1,2,\infty \}},
\end{equation}
in which
\begin{equation}
\begin{cases}
    \mathscr{I}_{sim} := \{\vec{x}\in \Omega \  \text{s.t.} \  \phi(\vec{x})\Bigl|_{t=t_f}=0.5\},\\
    \mathscr{I}_{true} := \{\vec{x}\in \Omega \  \text{s.t.} \  \phi(\vec{x})\Bigl|_{t=0}=0.5\}.\\
\end{cases}    
\end{equation}
Here $\Omega$ represents the domain and $\mathscr{I}_{sim},\mathscr{I}_{true}$ contain sets of points on the $\phi=0.5$ contour corresponding to the simulation and the ideal case respectively, as prescribed by \cite{nguyen2017volume}. Fig. \ref{fig:zalesak-overview} demonstrates the qualitative, anti-diffusive influence of the interface sharpening component on the Zalesak disk following a full rotation. Fig. \ref{fig:zalesak-contours} visualizes the sharpened Zalesak disk for grid spacings $h\sim 2\times 10^{-2},  h\sim 10^{-2}, $ and $ h\sim 5\times 10^{-3}$ as it rotates in $[0,t_f]$, and Fig. \ref{fig:zalesak-convergence} quantifies the interface retention according to the parameter defined in Eq.~\eqref{eq-def-Idelta} for the same cases.

In comparison to results reported by Nguyen \cite{nguyen2017volume}, the absolute error magnitude for \(h \sim 10^{-2}\) is larger; however this study reports higher convergence orders according to the following analysis. Nguyen presents convergence results as a function of Courant number $C$. Grid spacing exhibits the proportionality $h\propto 1/C$, implying that convergence orders derived from error variation with \(C\) can be mapped onto those derived from error variation with \(h\). According to this map, convergence orders of 0.71, 0.72, and 0.80 for \(\mathscr{I}^\delta_1\), \(\mathscr{I}^\delta_2\), and \(\mathscr{I}^\delta_\infty\), respectively, can be computed from \cite{nguyen2017volume}. Whereas the results in this study exhibit steeper convergence for \(\mathscr{I}^\delta_1\) and \(\mathscr{I}^\delta_2\) at 1.19 and 1.16. The convergence for \(\mathscr{I}^\delta_\infty\) is slightly shallower at 0.69. To enable a direct visual comparison, the slopes computed from \cite{nguyen2017volume} are plotted as reference lines in Fig. \ref{fig:zalesak-convergence}.

\color{black}

\subsubsection{Shape recovery}\label{sec-shape}

The purpose of this section is to measure the accuracy and conservation properties of the combined interface sharpening and surface tension models in 2D and 3D. The test cases in this section are motivated by $(i)$ the principle that surface tension acts to transform shapes with initially nonuniform curvature into an $n$-sphere (a circle in 2D or a sphere in 3D). This in addition to $(ii)$ the principle of mass conservation per phase implies that, given an initial analytical interface, a final uniform radius corresponding to the equilibrium position of surface tension-driven motion can be obtained. This section is divided into two parts, dedicated to proving the validity of $(i),(ii)$ within the proposed model in this study. Sec. \ref{sec-ly2d} uses the Young-Laplace condition to compute the steady-state interface size, shape, and pressure jump associated with surface tension-driven displacement of a four-pointed star, and compares this value to the simulated behavior. Sec. \ref{sec-ly3d} utilizes a 3D abstraction of the star shape  to simulate time-evolved mass loss per phase. 

\color{black}

  Sec. \ref{sec-ly2d} quantifies the extent to which the combined models are consistent with the Young-Laplace equation during a process of 

\paragraph{Young-Laplace consistency}\label{sec-ly2d}
  
As a way to demonstrate the capability of the model to compute precise body force terms even where highly positive-curvature and negative-curvature regions are in close proximity, the interface is assumed to take the shape of a four-pointed star. The initial radius of the star obeys 

\begin{equation}
    r_0(\theta ) = L\sqrt{ \frac{\cos \bigl( \frac{2\sin^{-1}k + m\pi }{2n}  \bigl)}{\cos \bigl( \frac{2\sin^{-1}(k\cos n\theta ) + m\pi }{2n}   \bigl)}  }
\end{equation}
such that $\phi_0(r_0)=0.5$, where $m=27.4$, $n=4$, $k=0.5, L=2.7971\times10^{-2}$, and $\theta\in [0,2\pi ]$. Let subscripts $_{0,f,\infty }$ refer to initial, final, and far-field states respectively; let fluid $II$ be interior to the star and let fluid $I$ occupy the far field. Then, assuming an initially isobaric domain such that $\Delta p_0 = 0$,  the expected pressure jump incurred by surface tension deforming the star $\Delta p_f$ can be deduced from conservation principles. By mass conservation
\begin{equation}\label{eq-rf2}
    \rho_f^{(II)} r_f^2 =   \frac{\rho^{(II)}_0}{\pi }A_{star}
\end{equation}
in which $A_{star}=\frac{1}{2}\int_0^{2\pi } r_0^2 (\theta)d\theta$. Note that if $\rho_f^{(II)}\approx \rho_0^{(II)}$ can be reasonably approximated then Eq.\eqref{eq-rf2} simplifies to $r_f=\sqrt{A_{star}/\pi}$. If not, then taking from the Young-Laplace condition and the equation of state, the final density of fluid $II$ can be solved via
\begin{multline}\label{eq-deltapf-sigmarf}
    \Delta p_f = 
    \frac{\sigma}{r_f}\\
    \Longrightarrow \rho_f^{(II)} = \Bigl[\frac{\sigma}{r_f} +(\gamma^{(I)}-1)\rho_\infty ^{(I)}\frac{C_p^{(I)}T}{\gamma^{(I)}} +p^{0(I) }- p^{0(II)}\Bigl]\frac{\gamma^{(II)}}{(\gamma^{(II)}-1)C_p^{(II)}T} \\ := \beta_1 + \frac{\beta_2}{r_f} .
\end{multline}
Substituting this into Eq.~\eqref{eq-rf2},
\begin{equation}\label{eq-rfquad}
    \beta_1r_f^2 +  \beta_2 r_f - \frac{\rho_0^{(II)}}{\pi }A_{star} = 0.
\end{equation}
With Eq.~\eqref{eq-rfquad} the final radius can be deduced from the initial radius and the initial thermodynamic properties for two general stiffened gases. In this case, take $\sigma=2000,\gamma^{(I)}=1.400,\gamma^{(II)}=1.932,\rho_0^{(I)}=1.16,\rho_0^{(II)}=998.23,C_p^{(I)}=1004.50,C_p^{(II)}=8095.08,p^{0(I)}=0, p^{0(II)}=1.1645\times10^{9}$. Then as phase $II$ is nearly incompressible under $M\ll1$, density variations due to surface tension-driven pressure change are negligible; therefore Eq.\eqref{eq-rf2} simplifies to $r_f\approx 2.00\times10^{-2}$. 
 
Then the expected pressure jump across the interface between the fluids at $t=t_f$ is solved as $\Delta p_f = \sigma /r_f \approx 1.00\times  10^{5}$. 

The purpose of the rest of this section is to compare this expected interfacial pressure jump to what is obtained by the simulation, $\Delta p_{sim}$. To do this, introduce the metric $[\Delta p_{sim} - \sigma \kappa_f ](\sigma\kappa_f)^{-1}$ to evaluate the percent error of the simulation with respect to the analytical solution across time. In addition, as surface tension is expected to transform the star shape into a uniform radius in $\theta$, the nondimensional metric $r^*$ in Eq.\eqref{eq-def-rstar}
is introduced to compare the magnitude of radial variation along the interfacial ($\phi=0.5$) contour with the grid spacing magnitude. In other words $r^*$ informs the number of grid spaces across which the radius varies along the surface; therefore $r^*\longrightarrow 0$ is expected in time. Remaining domain setup details are as follows: let $x,y\in [-0.035,0.035]$ with grid spacing magnitude $h\sim 10^{-3}$; apply zero-derivative boundary conditions $\partial \{\rho^{(I)}\phi,\rho,\rho\vec{u},\rho e,p,\phi \}^T/\partial n = 0$  to all four sides; and let $\Delta t=10^{-7},$ $t_f=2\Delta t\times10^4$.

\vspace{0.5cm}

\begin{figure}[ht]
    \centering
    \hspace*{2cm}
    \scalebox{0.75}{
    \begin{overpic}[width=\linewidth,tics=10]{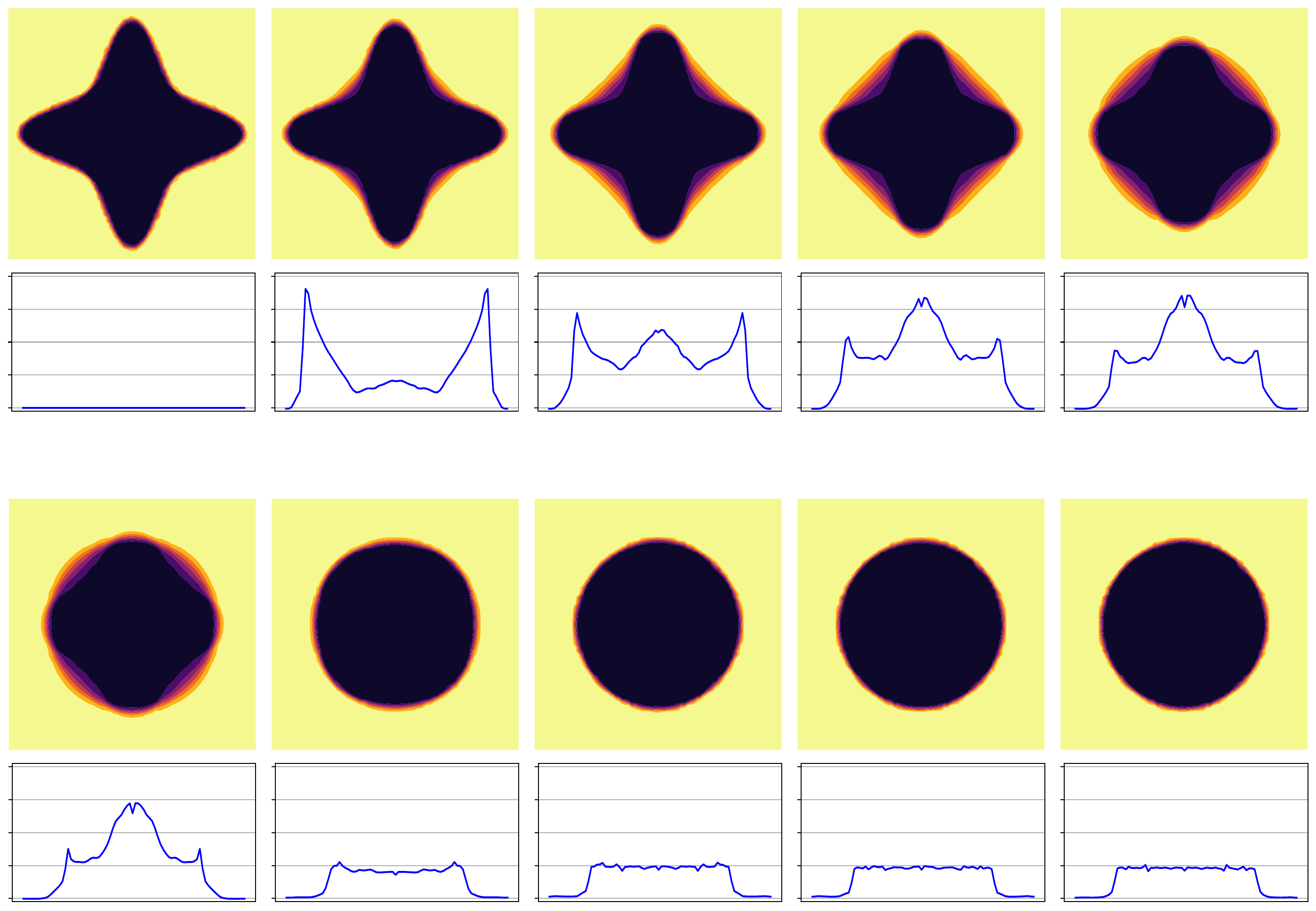}
            \put (-2,0) { $1$}
            \put (-2,2.5) { $2$}
            \put (-18,7) { $(b) \ p/10^5$}
            \put (-2,5) { $3$}
            \put (-2,7.5) { $4$}
            \put (-2,10) { $5$}

            \put (-2,37.5) { $1$}
            \put (-2,40) { $2$}
            \put (-18,44.5) { $(b) \ p/10^5$}
            \put (-2,42.5) { $3$}
            \put (-2,45) { $4$}
            \put (-2,47.5) { $5$}

            \put (9,13.5) { $\phi=1 $}
            \put (5,20) { \color{white}$\phi=0$\color{black}}

            \put (0,51) { $\phi=1 $}
            \put (5,57.5) { \color{white}$\phi=0$\color{black}}

            \put (-12,69.5) { transient: $t/t_f =0.00$}
            \put (20,69.5) { $t/t_f =0.04$}
            \put (40,69.5) { $t/t_f =0.08$}
            \put (60,69.5) { $t/t_f =0.12$}
            \put (80,69.5) { $t/t_f =0.16$}

            \put (0,32) {  $t/t_f =0.20$}
            \put (20,32) { $t/t_f =0.40$}
            \put (40,32) { $t/t_f =0.60$}
            \put (60,32) { $t/t_f =0.80$}
            \put (80,32) { $t/t_f =1.00$}

            \put (-17,57.5) {$(a) \ \phi$}
            \put (-17,20) {$(a) \ \phi$}

        \end{overpic}
        }
        \caption{Time evolution of a star-shaped interface relaxing to a circle under surface tension. $(a)$: $\phi$ field; $(b)$: Associated pressure jump at $y=0$ slice.}
        \label{fig-recover-star}
\end{figure}
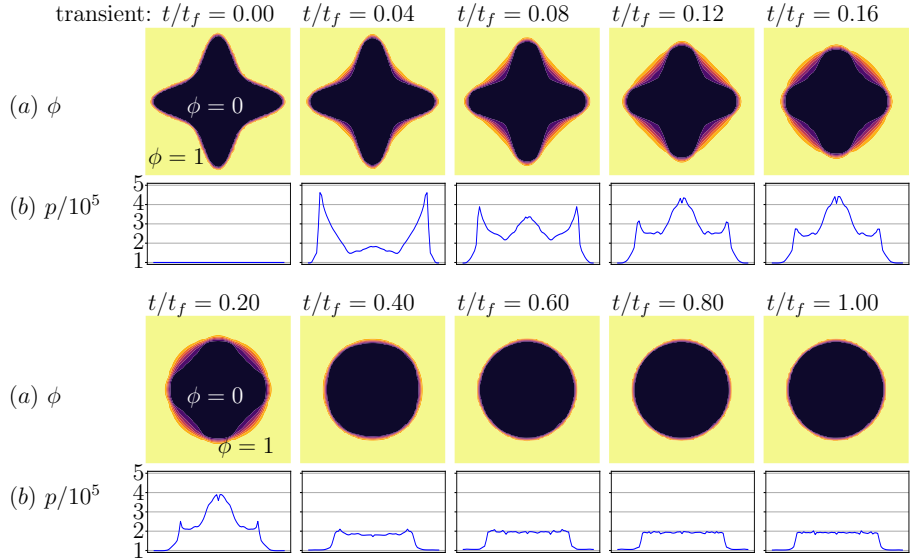

\begin{figure}[ht]
    \centering
        \makebox[\textwidth]{
        \includegraphics[width=\linewidth]{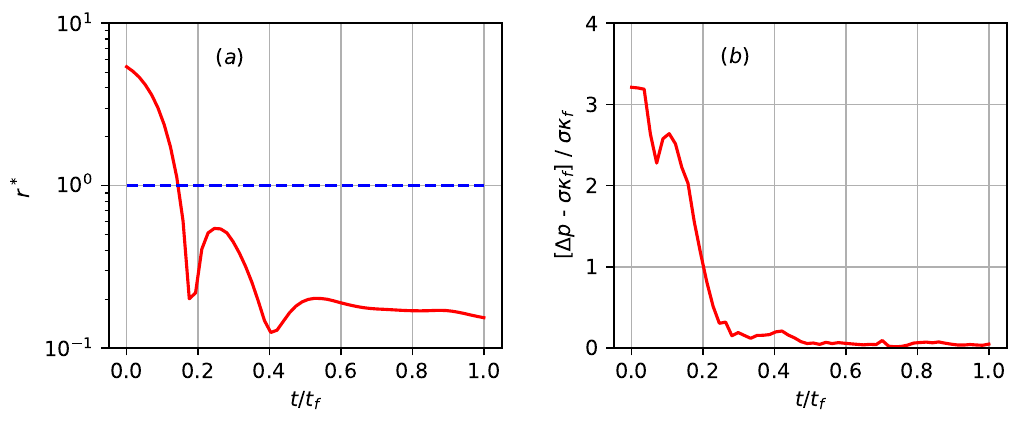}
    }
    \caption{$(a)$ Time-evolved normalized radial  standard deviation $r^*$. Blue line indicates $r^*=1$, i.e. threshold under which radius deviates less than one grid space magnitude.  $(b)$ Time-evolved Young-Laplace percent error $[\Delta p-\sigma \kappa_f]/\sigma\kappa_f \longrightarrow 0$ indicating convergence to expected solution.}
    \label{fig-star-combined}
\end{figure}

The time-evolved volume fraction field and simultaneous pressure cross-section at $y=0$ corresponding to this simulation is visualized in Fig. \ref{fig-recover-star}. In addition the time-evolved metrics $r^*,[\Delta p_{sim}-\sigma \kappa_f](\sigma\kappa_f)^{-1}$ are shown in Fig. \ref{fig-star-combined}. The interface initially oscillates in the radial direction as measured by $r^*(t/t_f \in [0,0.4])$ in Fig. \ref{fig-star-combined}$(a)$ because of the high energy state associated with the star shape. However, the interface quickly approaches a uniform circle as it oscillates, such that the standard radial deviation as $t\longrightarrow t_f$ is comparable to approximately $h/5$, as measured by $r^*(t/t_f=1)$ in Fig. \ref{fig-star-combined}$(a)$. In addition, notice the rapid convergence to a pressure jump of $\Delta p_{sim}\longrightarrow \sigma \kappa_f$ in Fig. \ref{fig-star-combined}$(b)$, aligning with  Eq.~\eqref{eq-deltapf-sigmarf}.

\FloatBarrier

\paragraph{Phase conservation}\label{sec-ly3d} 

The multiphase flow model assumes conservation of partial densities of both phases. In particular the mixture density, $\rho$, and the partial density of phase $I$, $\rho^{(I)}\phi$, are conserved, which imply the conservation of $\rho^{(II)}\phi^{(II)}$ by the condition $\phi^{(I)}+\phi^{(II)}=1$. As discussed in Sec. \ref{sec-shape}, this model assumption is physically accurate and underpins the derivation of the expected equilibrium position of a fluid body. Therefore it is of interest to quantify the effect of interface sharpening, introduced as an artificial RHS contribution applied to phase $I$, on phase mass conservation over time. In particular the presence of a high density ratio between the two phases causes small numerical errors in the volume fraction to produce disproportionately large phase density errors.

\begin{figure}[ht]
    \centering
    \begin{subfigure}[b]{\linewidth}
        \centering
        \includegraphics[width=\linewidth]{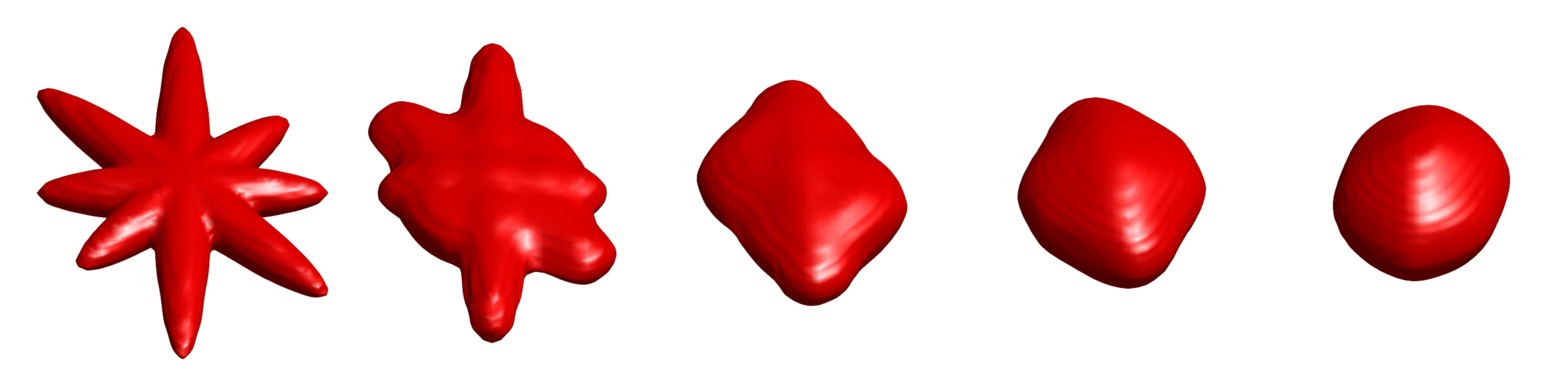}
        \caption{}
        \label{fig-ly3d-frames}
    \end{subfigure}
    \vskip\baselineskip
    \begin{subfigure}[b]{\linewidth}
        \centering
        \includegraphics[width=\linewidth]{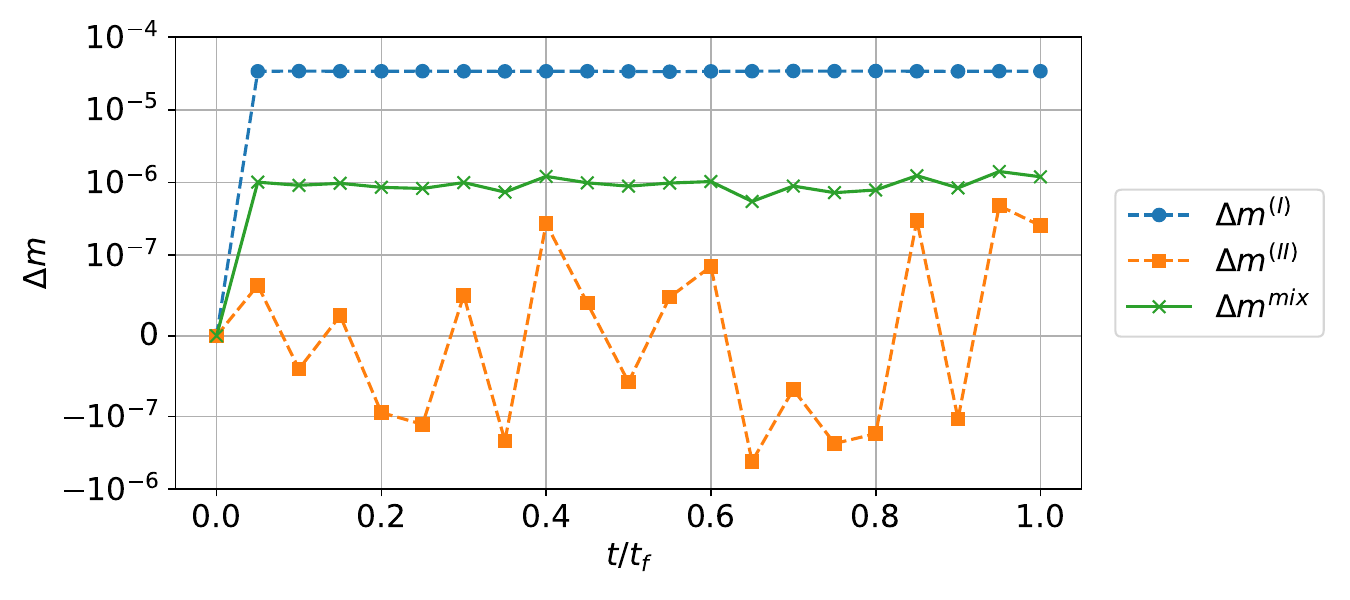}
        \caption{}
        \label{fig-ly3d-masscons}
    \end{subfigure}
    \caption{
    $(a)$ Interface evolution corresponding to 3D shape recovery test for $t\in [0,t_f]$, showing qualitative convergence to a sphere. $(b)$ Mass conservation assessment of the shape recovery test, showing the relative mass deviation \(\Delta m^k=f(t/t_f)\) for each phase and the mixture.}
    \label{fig-ly3d-combined}
\end{figure}

The desire to stress-test the model in this manner motivates the initial conditions $\gamma^{(I)}=1.400,\gamma^{(II)}=1.932,\rho_0^{(I)}=1.16,\rho_0^{(II)}=998.23,C_p^{(I)}=1004.50,C_p^{(II)}=8095.08,p^{0(I)}=0, p^{0(II)}=1.1645\times10^{9}$. The interface geometry used is a 3D abstraction of the surface tension-driven shape recovery presented in Sec. \ref{sec-ly2d}, though in this case constructed out of a set of conjoined ellipses instead of through a single radial function. Remaining domain setup details are as follows: let $x,y,z\in [-0.035,0.035]$ with grid spacing magnitude $h\sim 10^{-3}$; apply zero-derivative boundary conditions $\partial \{\rho^{(I)}\phi,\rho,\rho\vec{u},\rho e,p,\phi \}^T/\partial n = 0$  to all sides; and let $\Delta t=10^{-7},$ $t_f=2\times10^{-3}$. Fig. \ref{fig-ly3d-frames} shows the time-evolved interface evolution, qualitatively matching the expectation of convergence to uniform curvature. Fig. \ref{fig-ly3d-masscons} is a visualization of the time-evolved mass change in the domain, measured by the parameter
\begin{equation}\label{eq-mass-deviation-discrete}
    \Delta m^k(t/t_f) := \frac{\sum_i \left( \phi^k \rho^k \right)_i\Bigl|_{t/t_f\in [0,1]}}{\sum_i \left( \phi^k \rho^k \right)_i\Bigl|_{t=0}} - 1, \quad 
    k \in \{(I), (II), \mathrm{mix}. \}
\end{equation}
which is used to track both phases as well as the mixture. As shown, the interface sharpening term disproportionately distorts the mass of phase $I$. This is expected as the lighter fluid in the system incurs greater percent deviation in response to numerical changes in the volume fraction. However even as this case is designed to produce the worst case scenario, the order of magnitude of $\Delta m^k \ \forall \  k $ can be considered negligible.

\begin{figure}[ht]
    \centering
    \includegraphics[width=\linewidth]{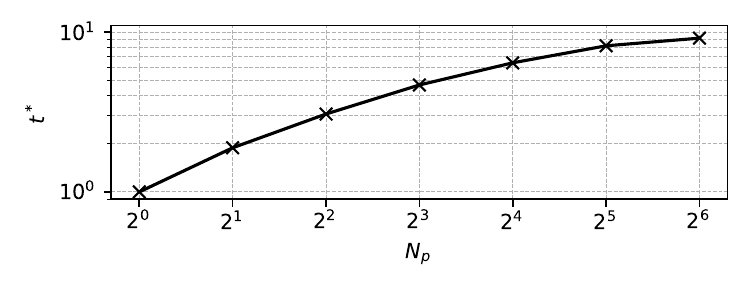}
    \caption{
    Strong scaling performance in the 3D shape recovery test.}
    \label{fig-ly3d-scalability}
\end{figure}

\paragraph{Scalability} Strong scaling characterizes how solve time $t_{sim}$ scales with the number of processing cores $N_p$. The dimensionless speedup metric $t^* = t_{sim}(1)/t_{sim}(N_p)$ is chosen to quantify the reduction in simulation time for $N_p$ cores relative to a single core. In this context a speedup slope of 1 corresponds to a perfect parallel efficiency where the doubling of cores results in a halving of runtime. In contrast, the case presented in Sec. \ref{sec-ly3d} was run for $N_p=2^k$, $k\in [0,6]$ with a fitted $t^*$ slope of 0.53, as shown in Fig. \ref{fig-ly3d-scalability}. This indicates sublinear but still significant strong scaling efficiency over the tested range.

\color{black}

\subsection{Droplet pinchoff}\label{sec-droplet}

When simulating multiphase flows where surface tension causes one of the phases to split into two or more components, the simulation's ability to accurately capture this behavior relies heavily on capturing the interface location. Therefore, the combination of interface sharpening and surface tension is crucial towards systems including droplet pinchoff. A widely studied problem is the deformation of a primary phase induced by the shear velocity of a secondary phase. This shearing causes wavelike surface instabilities, and curvature-driven surface tension punches these waves at their necks so that discrete droplets of the primary phase are detached from the larger body and entrained with the secondary phase. For systems exhibiting this behavior, a model is required that accurately captures the shapes and sizes of droplets based on both the physical properties of the two phases and  the flow characteristics.

\begin{figure}[ht]
    \centering
    \includegraphics[width=0.6\linewidth]{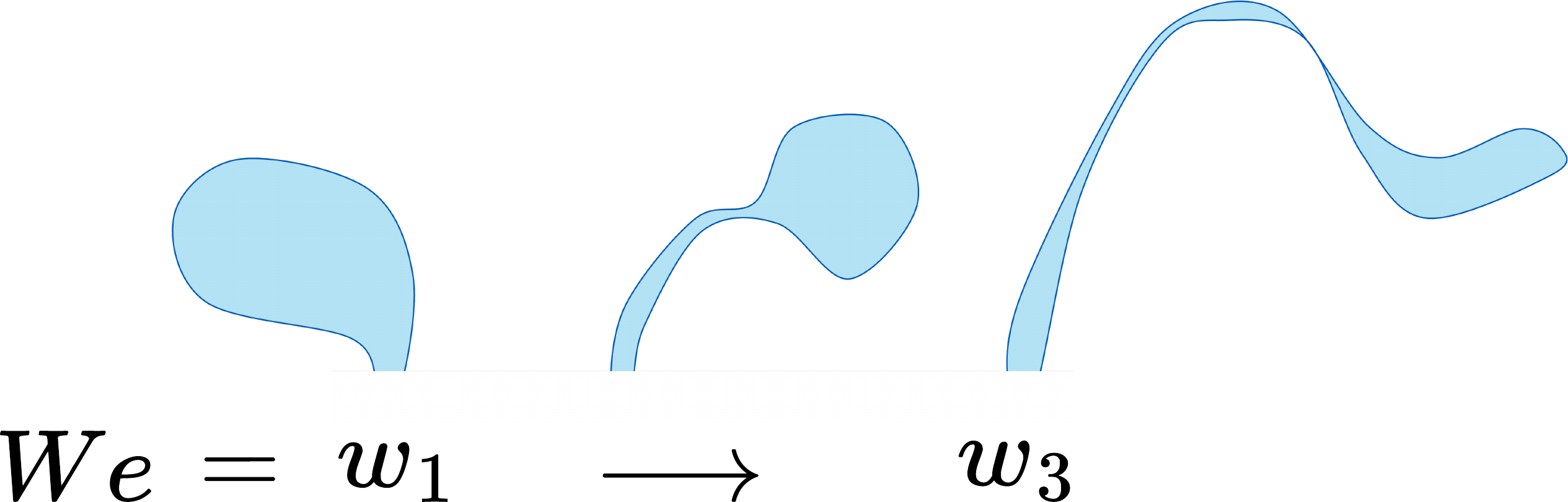}
    \caption{Typical droplet shapes at the onset of pinchoff based on $We,Oh$ regime.}
    \label{fig-jain-droplet}
\end{figure}

The dynamics of droplet breakup are well-understood as a function of Weber number ($We=\rho^{*}||\vec{u}_0||_{max}^{2}r_0/\sigma $) 
and Ohnesorge number ($Oh=\sqrt{We}/Re$, $Re=\rho^*||\vec{u}_0||_{max}r_0/\mu^*$), which encapsulate the relative contributions of viscosity (ratio $\mu^*=\mu^{(I)}/\mu^{(II)}$), kinetic energy density (with density ratio $\rho^* =\rho^{(I)}/\rho^{(II)}$), and surface tension ($\sigma$) as the primary and secondary phases interact. Schmehl et al generate a set of polynomials $w_i = f(Oh)$ that outline various droplet breakup regimes \cite{schmehl2000cfd}. They are
\begin{equation}\label{eq-breakupregime2}
    \begin{cases}
        We<12(1+1.077 Oh^{1.6}) := w_1, & \text{deformation},\\
        w_1<We<20(1+1.2 Oh^{1.5}) := w_2, & \text{bag breakup},\\
        w_2<We<32(1+1.5 Oh^{1.4}) := w_3, & \text{multimodal breakup},\\
        w_3<We, & \text{shear breakup},\\
    \end{cases}
\end{equation}
where 2D slices of the above droplet behaviors, adapted from the literature \cite{jain2019secondary}, are visualized in Fig. \ref{fig-jain-droplet}.

At 
$We\sim  w_1$\color{black}, surface area minimization of the primary phase far exceeds the energy required to elongate from velocity shear. Around this regime, the droplet oscillates at a natural frequency while converging to a spherical shape. As the Weber number is set in the direction of $w_1\longrightarrow w_3$, the flow becomes decreasingly surface tension-driven and increasingly kinetics-driven. As this happens, the droplet transforms into a flat, stringlike shape. For all nonzero, finite Weber numbers, the primary phase will to some extent form a rounded tip held together by surface tension and will also to some extent incur stringlike elongation caused by kinetic energy. These components are conventionally labeled as the \textit{child droplet} and the \textit{parent} respectively. In this study, the relative sizes of these components are quantified to match the range of $We,Oh$ combinations present in Sementilli \cite{sementilli2022HRM,budzinski2020radiation}, whose inputs lead to $Oh\sim \mathcal{O}(10^{-3}), \ \ We\sim \mathcal{O}(10^1).$
 
However, the model is generalizable to arbitrary $We,Oh$ numbers.

\begin{figure}[ht]
    \centering
    \scalebox{1}{
    \begin{overpic}[width=\linewidth]{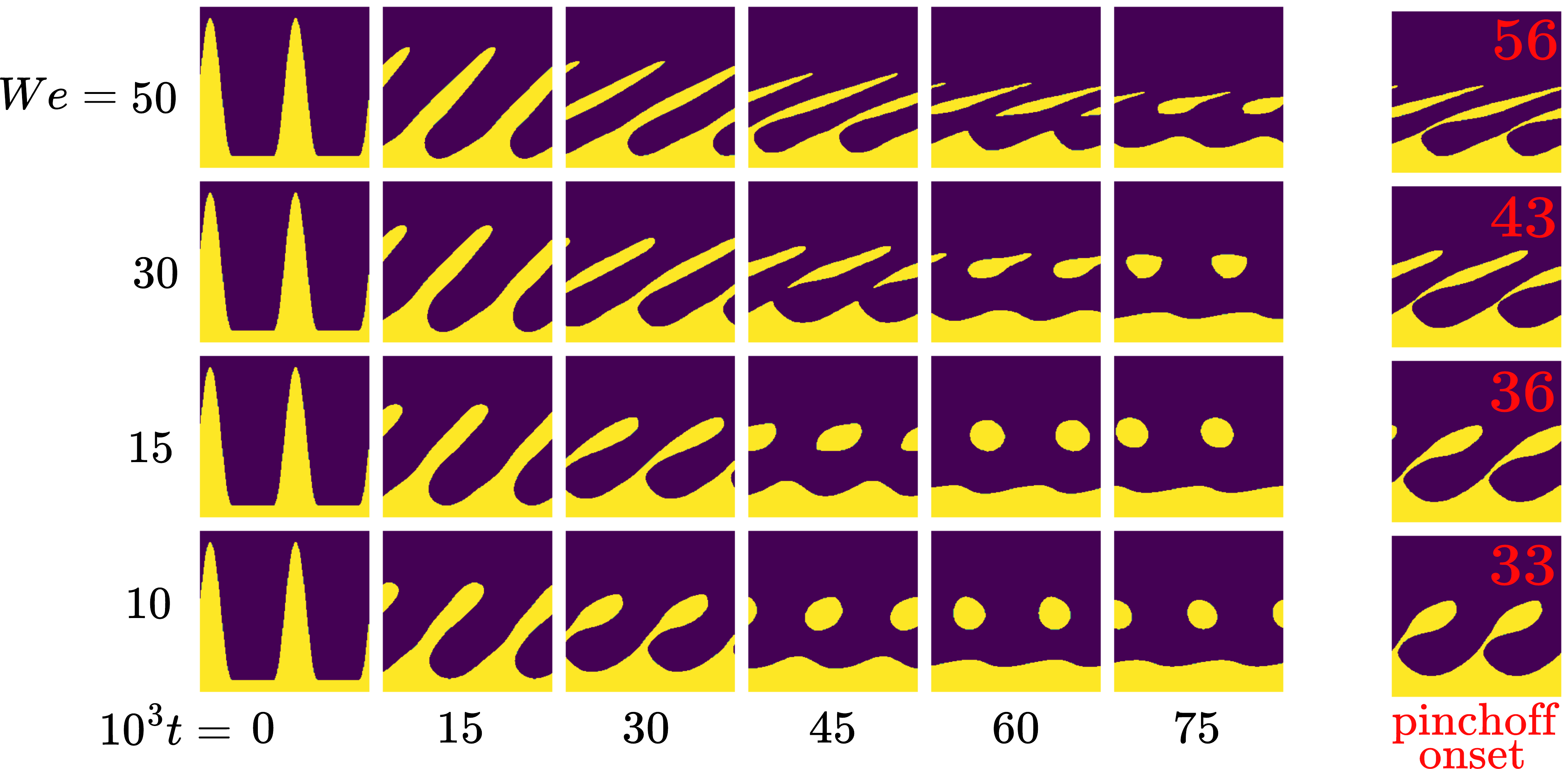}

    \end{overpic}
    }
    \caption{Time-evolved droplet dynamics at $t=0,15,30,45,60,75\times 10^{-3}$, as well as the times of pinchoff onset across various Weber numbers.}
    \label{fig-droplet-map}
\end{figure}

The remainder of this study focuses on comparing several multiphase simulations within these orders of magnitude to the expected droplet behavior offered by Eq. \ref{eq-breakupregime2}. As a way to elicit various droplet shapes,
 
a case is devised wherein the primary phase has highly fluctuant curvature and a transverse mass flux of the secondary phase is imposed. The details of the case are as follows. The domain is bounded by $x,y\in [0,0.2]=[0,r_0]$ and the grid spacing magnitude is $h\sim 0.001$ with quadrilateral elements. The initial location of the interface between the two phases is determined by
\begin{equation*}
    f(x) = g(x)h(x) + \frac{y_{max}(1-A)}{2},
\end{equation*}
\begin{equation}
    g(x) = \frac{1}{2}Ay_{max}\Bigl(1 + \sin \Bigl[ \frac{2\pi x}{T_f}  \Bigl]\Bigl), \ \  h(x) = \frac{1}{2}\Bigl( 1 + \text{sign}\Bigl[ \sin \Bigl\{ \frac{\pi x}{T_f} + \frac{\pi }{4}   \Bigl\}   \Bigl]   \Bigl),
\end{equation}
where parameters are $T_f=x_{max}/(2N)$, $N=4,$  and $A=0.85$. The bottom boundary is set as no-slip such that $||\vec{u}||=0$ and the top boundary is set as zero-gradient  such that $\partial \{\rho^{(I)}\phi,\rho,\rho\vec{u},\rho e,p,\phi \}^T/\partial y  = 0$. The left and right sides are set as periodic to mimic a horizontally infinite domain. Elsewhere in the domain the velocity field is initially set as $\vec{u}_0=\{\min (100,\max[100(y-0.015)/0.17,0]), \ 0 \}^T\Rightarrow ||\vec{u}_0||_{max}=100$ to represent secondary phase shearing. The time step is set as $\Delta t =10^{-7}$. Finally Table \ref{tab-weoh} shows the various $We,Oh$ combinations tested in this study. The results of this set of tests are visualized in Fig. \ref{fig-droplet-map}.

\begin{table}[ht]
\centering
\begin{tabular}{|l|l|l|l|l|l|l|}
\hline
$We$ & $Oh$      & Droplet shape description & $\rho^*$ & $10^3\mu^*$ & $\sigma $  \\ \hline
10   & $10^{-3}$ & Deformation  & $1.05$ & $6.641$ & $210$               \\ \hline
15   & $10^{-3}$ & Bag breakup  & $1.05$ & $5.422$ & $140$        \\ \hline
30   & $10^{-3}$ & Multimodal breakup  & $1.05$ & $3.834$ & $70$    \\ \hline
50   & $10^{-3}$ & Shear breakup  & $1.05$ & $2.970$ & $42$          \\ \hline
\end{tabular}
\caption{Weber and Ohnesorge numbers used in this study and associated constants.}

\label{tab-weoh}
\end{table}

\begin{figure}[ht]
    \centering
        \makebox[\textwidth]{
        \includegraphics[width=\linewidth]{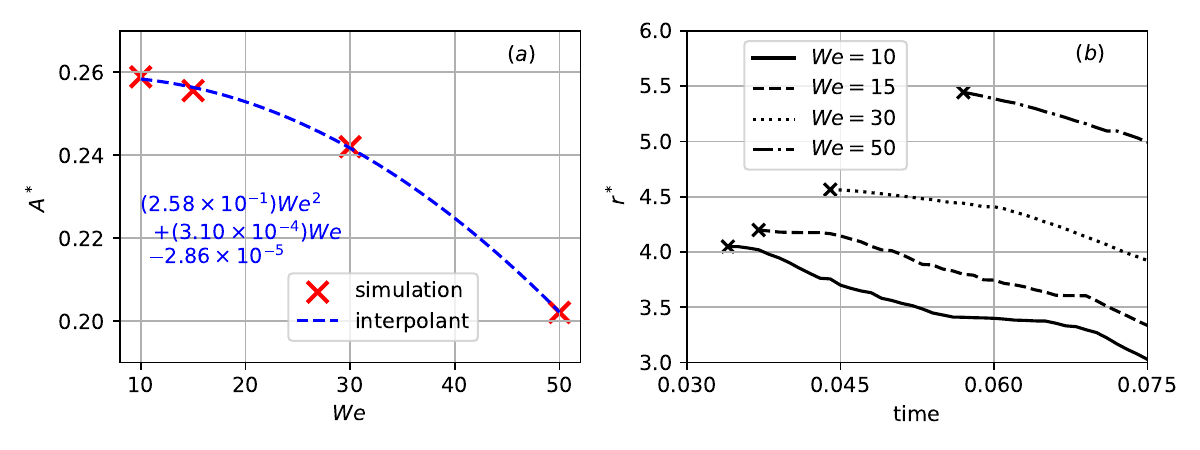}
    }
    \caption{$(a)$ Child droplet area ratio at time of pinchoff onset across various Weber numbers. Blue line indicates interpolant to $R^2=0.999$. $(b)$ Time-evolved normalized standard deviation of child droplet radius across various Weber numbers.}
    \label{fig-area-child}
\end{figure}

 At the highest Weber number the flow is predominated by kinetic effects; therefore, pinchoff onset is delayed and the primary phase is allowed to elongate for a substantially greater amount of time. At the pinchoff onset, the child droplet subsumes a much lesser area than the parent, as shown by Fig. \ref{fig-jain-droplet}. However, as the Weber number decreases such that surface tension contributes more significantly to the flow characteristics, pinchoff onset is more rapid. This is because the child droplet converges to a uniform curvature more rapidly, which creates a highly curvaceous region between the bottom of the child droplet and the parent. Because the necking speed increases, the location of pinchoff is lower towards the bottom of the domain and therefore the size of the child droplet relative to the parent is larger, as illustrated by Fig. \ref{fig-area-child}$(a)$, wherein the child droplet's area ratio is $A^* = A_{child}/(A_{child}+A_{parent})$. 
 This metric represents a normalized mass entrainment magnitude, which is of concern in the context of rocket motors and fluid injectors \cite{klyus2024effect,khakimov2024improving}, among other applications. In addition, the rate that the child droplet reforms to a uniform curvature, or minimizes its surface area to volume ratio, is also sought as a metric for mixing efficiency. Therefore, a quantity of interest is the normalized standard deviation of the interfacial radius, $r^*$ in Eq. \ref{eq-def-rstar}. Fig. \ref{fig-area-child}$(b)$ plots the time evolved diminishing of the radial deviation starting at the moment of pinchoff. Across Weber numbers the slope of the convergence towards $r^*\longrightarrow 0$ is shown to be roughly uniform, but the initial deviation is informed by the strength of surface tension, which is expected because highly elliptical drops contain both high- and low-curvature regions simultaneously.
\FloatBarrier

\section{Conclusion}\label{sec-conclusion}

Numerical diffusion inherent to discrete approximations of PDE systems manifests as interface smearing in the context of multiphase flows. To mitigate this effect, this study proposes an interface sharpening procedure for a compressible multiphase flow system with surface tension casted onto an arbitrarily constructed grid. As a means of validation, the model is shown to accurately satisfy the Young-Laplace condition, where a high-pressure region is bounded by surface forces across an accurately located sharp interface. In addition a star with oscillatory  surface curvature is shown to converge to a sharpened circle, as surface tension and interface sharpening modules are combined. As a physical application of these modules, an interfacial geometry was devised to elicit various droplet shapes based on the Weber and Ohnesorge numbers characterizing the shearing flow. The  curvature deviation across the interface of the child droplet decreases with a lower Weber number, and the child droplet area fraction is shown to be directly proportional to the necking speed. These results are consistent with the surrounding literature; therefore the model offered in this study is a precise tool in evaluating the interfacial dynamics of diverse kinds of multiphase compressible flows on an arbitrarily constructed grid.
\section{Acknowledgements}\label{sec-acknowledgements}

This research is funded by the United States Department of Energy’s (DoE) National Nuclear Security Administration (NNSA) under the Predictive Science Academic Alliance Program III (PSAAP III) at the University at Buffalo, under contract number DE-NA0003961. Support was provided by the Center for Computational Research at The University at Buffalo. The authors are very grateful and wish to thank Matthew McGurn of the University at Buffalo for his generous spirit and helpful conversations.
\appendix

\section{Gaussian convolution}\label{sec-convolve}

The Gaussian distribution function introduced in Sec. \ref{sec-mfs} is written as 
\begin{equation}
    G(x) = \frac{1}{\sqrt{2\pi }s}e^{-x^2/(2s^2)}
\end{equation}
where $s$ is the standard deviation. Trivially $\int_{-\infty}^\infty G(x)dx=1$ is a convolving function that would result in no loss in the total area of the field that is being convolved. However, from a computational perspective it becomes necessary to bound this integral finitely. It then becomes important to understand the relationship between the magnitude of the integral bounds and the amount of area under the curve that is lost due to convolution. If the bounds are set as a multiple $N$ of the standard deviation, then 
\begin{equation}
    \int_{-Ns}^{Ns}G(x)dx = 1-10^{-m}
\end{equation}
is posed to quantify the relationship between the number of standard deviations one integrates along and the order of magnitude of area loss, $m$. This equation is simplified to
\begin{equation}\label{eq-nofm}
    N(m) = \sqrt{2}\text{erf}^{-1}(1-10^{-m}),
\end{equation}
and this is useful when a ceiling for error has been predetermined. In this study, the preservation of 99\% of the area under the curve is prescribed, which corresponds to $m=2 \rightarrow N\approx 2.6$. (Preserving $99.9\%$ of the area requires $m=3 \rightarrow N\approx 3.3$, etc.) 

The range of the smoothing is  quantified by the magnitude of the standard deviation $s$. Let $s=Ch$, where $h$ is the grid spacing length scale, and $C$ is a constant (in this study, $C=1$ is selected). Then the number of cell layers that the convolution occurs across, $\mathscr{L}$, must obey the relation \begin{equation}\label{eq-lcn}
    \mathscr{L} = \ceil[\big]{\frac{Ns}{h}} = \ceil[\big]{CN},
\end{equation}
where the top-heavy brackets denote a ceiling operator.

After choosing the standard deviation magnitude ($C$ such that $s=Ch$) and the number of standard deviations that the convolution is bounded by ($N$ in Eq. \ref{eq-nofm}), the number of cell layers involved in smoothing is computed via Eq. \ref{eq-lcn}. This is useful because a mask of $\mathscr{L}$ layers about the volume fraction interface can be identified prior to smoothing. Then the Gaussian smoothing algorithm only attributes nonzero values to cells that are contained by the mask. This prevents computation of interface sharpening terms on cells that are sufficiently distant from the interface.
\bibliographystyle{elsarticle-num} 

\bibliography{7-0bib}

\end{document}